\documentclass[useAMS,usenatbib]{mn2e}
\usepackage{amssymb}
\usepackage{lscape}
\usepackage[dvips]{graphicx}
\newcommand{\etal}{\hbox{et al.}} 
\newcommand{\vlsr}{\hbox{v$_{\textrm{\tiny{LSR}}}$}}
\newcommand{\Trms}{\hbox{T$_{\textrm{\tiny{RMS}}}$}}
\newcommand{\fwhm}{\hbox{$\Delta$v$_{\textrm{\tiny{FWHM}}}$}}
\newcommand{\Tpeak}{\hbox{T$_{\textrm{\tiny{Peak}}}$}}
\newcommand{\Trot}{\hbox{T$_{\textrm{\tiny{rot}}}$}}
\newcommand{\Nh}{\hbox{$\overline{N_{H_2}}$}}
\newcommand{\nh}{\hbox{$\overline{n_{H_2}}$}}
\newcommand{\nH}{\hbox{$n_{H_2}$}}

\title[Interstellar gas towards CTB\,37A]{Interstellar gas towards CTB\,37A and the TeV gamma-ray source HESS\,J1714$-$385}
\author[Nigel I. Maxted \etal]{Nigel I. Maxted$^{1}$\thanks{E-mail: nigel.maxted@adelaide.edu.au}, Gavin P. Rowell$^{1}$, Bruce R. Dawson$^{1}$, Michael G. Burton$^{2}$, \newauthor Yasuo Fukui$^{3}$, Andrew Walsh$^{4}$, Akiko Kawamura$^{3}$, Hirotaka Horachi$^{3}$, \newauthor Hidetoshi Sano$^{3}$, Satoshi Yoshiike$^{3}$ \& Tatsuya Fukuda$^{3}$\\
$^{1}$School of Chemistry \& Physics, University of Adelaide, Adelaide, 5005,  Australia\\
$^{2}$School of Physics, University of New South Wales, Sydney, 2052, Australia\\
$^{3}$School of Physics, Nagoya University, Furocho, Chikusa-ku, Nagoya, Aichi, 464-8602, Japan\\
$^{4}$International Centre for Radio Astronomy Research, Curtin University, GPO Box U1987, Perth, Australia\\ 
}
\begin{document}

\date{Accepted 2013 June 23. Received 2013 June 20; in original form 2012 October 26}

\maketitle

\label{firstpage}

\begin{abstract}
Observations of dense molecular gas towards the supernova remnants CTB\,37A (G348.5$+$0.1) and G348.5$-$0.0 were carried out using the Mopra and Nanten2 radio telescopes. We present CO(2-1) and CS(1-0) emission maps of a region encompassing the CTB\,37A TeV gamma-ray emission, HESS\,J1714$-$385, revealing regions of dense gas within associated molecular clouds. Some gas displays good overlap with gamma-ray emission, consistent with hadronic gamma-ray emission scenarios. 
Masses of gas towards the HESS\,J1714$-$385 TeV gamma-ray emission region were estimated, and were of the order of 10$^3$-10$^4$\,M$_{\odot}$. In the case of a purely hadronic origin for the gamma ray emission, the cosmic ray flux enhancement is $\sim$80-1100 times the local solar value. This enhancement factor and other considerations allow a discussion of the age of CTB\,37A, which is consistent with $\sim$10$^4$\,yr.

\end{abstract}


\section{Introduction}\label{sec:intro}
Gamma ray observations may be the key to solving one the of longest unsolved mysteries in astrophysics, the origin of cosmic rays (hereafter CRs). Since gamma-rays are byproducts of cosmic ray interactions, high gamma-ray fluxes are expected from regions with enhanced cosmic ray densities, such as near cosmic ray acceleration sites. The leading theory to explain the acceleration of these particles is first order Fermi acceleration in the shocks of supernova remnants (hereafter SNRs).

A candidate region for CR-acceleration is CTB\,37, which is comprised of three SNRs: CTB\,37A (G348.5$+$0.1), CTB\,37B (G348.7$+$0.3) and G348.5$-$0.0. CTB\,37A was initially discovered at radio wavelengths \citep{Clark:1975} and was recently observed to emit at TeV energies (HESS\,J1714$-$385, \citeauthor{Aharonian:2008}, \citeyear{Aharonian:2008}). GeV-energy excesses have also been detected (3EG\,J1714$-$3857, \citeauthor{Green:1999}, \citeyear{Green:1999}; \citeauthor{Hartman:1999}, \citeyear{Hartman:1999}, 1FGL\,J1714.5$-$3830, \citeauthor{Castro:2010}, \citeyear{Castro:2010}).


\begin{figure*}
\centering 
\includegraphics[width=1.01\textwidth]{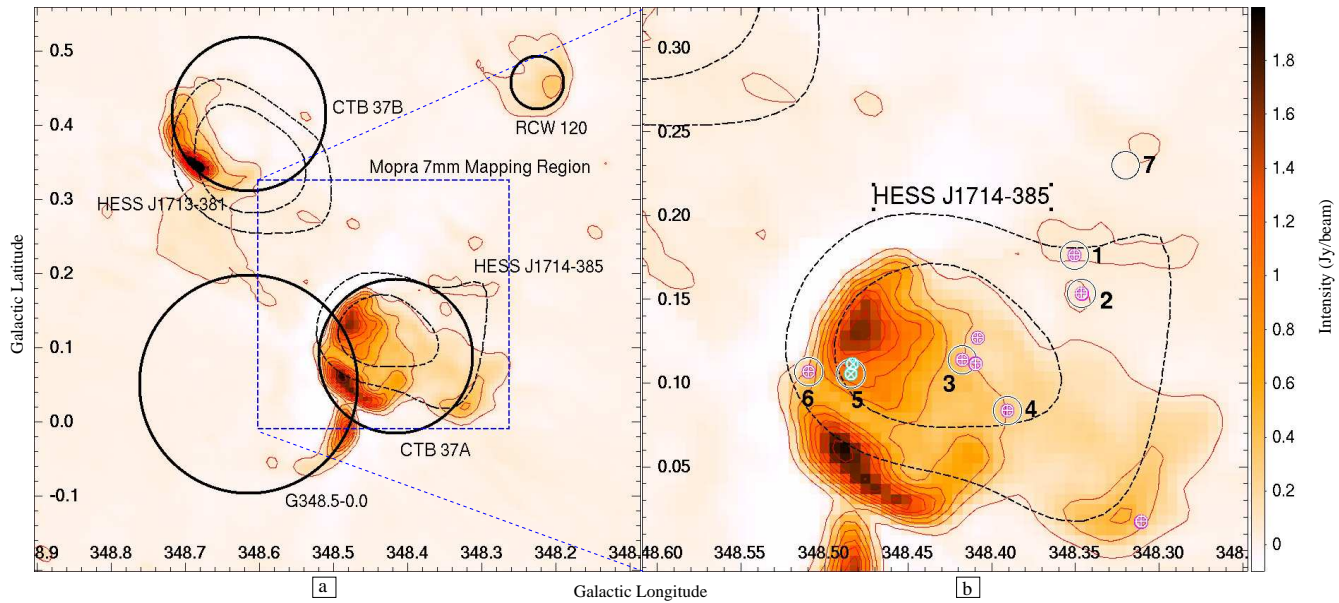}
\caption{\textbf{(a)} Molonglo 843\,MHz radio continuum image showing CTB\,37 \citep{Clark:1975}. 80 and 100 gamma-ray excess count contours of HESS\,J1714$-$385 (CTB\,37A) and HESS\,J1713$-$381 (CTB\,37B) (black, dashed) \protect\citep{Aharonian:2008} and Molonglo 843\,MHz radio continuum (bandwidth $\sim$4.4\,MHz) contours (red) are overlaid. A dashed blue square represents the region observed with Mopra (this work) and large labeled black circles mark the approximate locations and extents of objects of interest, as visually approximated from the 843\,MHz image. \textbf{(b)} Similar to (a), but zoomed on CTB\,37A. Crosses represent both the $\sim-$65\,km\,s$^{-1}$ (pink,$+$) and $\sim-$25\,km\,s$^{-1}$ (cyan,$\times$) OH maser populations \protect\citep{Frail:1996}, circles represent the position and beam FWHM of 7\,mm Mopra deep ON-OFF switched pointings. \label{fig:CTB1}}
\end{figure*}

The CTB\,37A radio continuum emission (see Figure\,\ref{fig:CTB1}a) is suggestive of a `break-out' morphology. Strong shell-shaped emission emanates from the eastern (galactic coordinate system) section and a weaker lobe or `break-away' region extends in a western direction. \citet{Reynoso:2000} observed CO(1-0) emission towards CTB\,37A and investigated the possibility that the observed features are caused by constraining molecular gas along the north, east and south boundaries. They found associated \vlsr$\sim$\,$-$65\,km\,s$^{-1}$ molecular clouds around the northern and eastern edges of the CTB\,37A radio continuum emission, while minimal gas was seen in the location of the `break-out' features to the west. In the south however, inconsistent with predictions, no gas (as traced by CO(1-0)) was seen to be constraining the remnant.

Knowledge of gas distribution is also important when considering a hadronic (pion production via CR collisions with gas) origin for the CTB\,37A gamma-ray emission, although the nature of the parent particle population (leptonic and/or hadronic) is not entirely clear. Non-thermal X-ray emission is absent from the CTB\,37A radio rim \citep{Frail:1996}, which, in the presence of strong magnetic fields would imply a lack of ongoing high energy electron acceleration. Indeed, \citet{Brogan:2000} measured strong magnetic fields (0.22-1.5\,mG) towards shocked regions of CTB\,37A, and in such environments TeV leptons have limited lifetimes due to synchrotron losses (synchrotron lifetimes of $\lesssim$250\,yr, very short compared to the SNR age, see \S\,\ref{sec:Age}). If the high energy lepton population is dimished towards the CTB\,37A radio rim, which it appears to be, a hadronic scenario for gamma-ray emission production would be favourable \citep{Aharonian:2008}.

Alternatively, a leptonic scenario for CTB\,37A is supported by the presence of a coincident region of non-thermal X-ray emission \citep{Aharonian:2008}. This could possibly indicate a pulsar wind nebulae (PWN) accelerating electrons within a blast wave (making CTB\,37A a composite SNR), but this hypothesis is complicated by the non-detection of a point-like source or pulsar within the extended non-thermal emission \citep{Aharonian:2008}. Certainly, the non-thermal and thermal emission from the outer rim and centre of CTB\,37A, respectively, indicate that CTB\,37A is a mixed-morphology SNR. Results by \citet{Brandt:2011} are suggestive of a significant Bremsstrahlung emission component between 1 and 100\,GeV, with a hadronic component dominating at TeV energies and above, but the question of the nature of the emitting particle population is still largely an open question.





Also present towards the CTB\,37A region \citep{Aharonian:2008}, are 1720\,MHz OH masers, which are indicators of shock-gas interactions, likely triggered by SNR shocks \citep{Aharonian:2008,Frail:1996}. However, the interpretation of this data is confused by the existence of a coincident partial shell SNR, G348.5$-$0.0 \citep{Milne:1979,Kassim:1991}, which can be seen in 843\,MHz radio continuum emission in Figure\,\ref{fig:CTB1}a. Two masers comprise a population at line-of-sight-velocity, \vlsr$\sim$\,$-$22\,km\,s$^{-1}$ while eight comprise a second population at \vlsr$\sim$\,$-$65\,km\,s$^{-1}$. Although it has been suggested that the 2 maser populations may be attributable to each of the 2 SNRs separately, recent work by \citet{Tian:2012} (hereafter TL2012) demonstrates that CTB\,37A may simply be associated with multiple varied-velocity clouds. This suggests that both populations of masers may be attributable to CTB\,37A (and neither would be associated with G348.5$-$0.0).

\citet{Reynoso:2000} found molecular CO(1-0) cloud associations towards most of the masers, but two \vlsr$\sim$\,$-$65\,km\,s$^{-1}$ OH masers (Locations 1 and 2, see Figure\,\ref{fig:CTB1}b) did not have clear CO emission counterparts at corresponding velocities and locations. The formation of 1720\,MHz OH masers requires large gas densities of \nH$\sim$10$^4$-10$^5$\,cm$^{-3}$ \citep{Draine:1983,Frail:1998}, but CO(1-0) emission has a critical density of $\lesssim$10$^3$\,cm$^{-3}$ and quickly becomes optically thick. As \citet{Reynoso:2000} suggest, to probe deeper, observations of dense gas tracers are required. 


 


In addition to being conducive to OH maser generation, dense gas may significantly affect the dynamics of the SNR shock motion, while also providing denser targets for gamma-ray-producing hadronic cosmic ray interactions. Knowing the location of dense gas may lead to both an explanation for the `break-out' morphology of the radio continuum emission and more detailed modeling of the CTB\,37A TeV emission. Previous studies have noted that dense gas may hinder CR diffusion and change the resultant high energy photon spectrum from hadronic interactions \citep{Gabici:2009,Casanova:2011,Fukui:2012,Maxted:2012}.

To complement existing Nanten2 CO(2-1) and Southern Galactic Plane Survey (SGPS) HI data \citep{McClure:2005,Haverkorn:2006,Tian:2012}  
we took 7\,mm observations of the CTB\,37A region with the Mopra radio telescope. We targeted the gas tracer CS(1-0) (critical density\,$\sim$8$\times$10$^4$\,cm$^{-3}$) to investigate dense gas 
and simultaneously searched for the potential shock-tracers SiO(1-0) and CH$_3$OH (motivated by results towards the SNR W28, \citeauthor{Nicholas7mm:2012}, \citeyear{Nicholas7mm:2012}) to search for shocked gas related to SNR activity. 

To investigate the gas distribution towards CTB\,37A, we applied gas parameter calculation methods presented in \S\ref{sec:GasParam} to several clouds of gas, which we consider separately in subsections of \S \ref{sec:GasDist}. In \S\ref{sec:X-ray} we compare column densities estimated by us to estimates by previous authors. In \S\ref{sec:tev} we use our gas mass estimates to estimate the cosmic ray density in several scenarios of gamma-ray emission from hadronic processes for CTB\,37A. Finally, in \S\ref{sec:Age}, we discuss the age of CTB\,37A.

\subsection{Distance}\label{sec:Distance}
By noting that towards CTB\,37A, HI absorption occurs in gas at \vlsr$\sim$\,$-$110\,km\,s$^{-1}$, but not from gas that lies at \vlsr$\sim$\,$-$65\,km\,s$^{-1}$, \citet{Caswell:1975} concluded that the \vlsr$\sim$\,$-$110\,km\,s$^{-1}$ gas was nearer to us than CTB\,37A. Assuming that the line-of-sight-velocities were primarily due to galactic kinematic motions, the authors recognised that the \vlsr$\sim$\,$-$65\,km\,s$^{-1}$ gas must be on the `far-side' (past the tangent point) of the galaxy in order to be spatially behind the \vlsr$\sim$\,$-$110\,km\,s$^{-1}$ gas. The distance of the CTB\,37A radio emission was thus constrained to be between 6.7 and 13.7\,kpc (assuming l$\sim$349$^{\circ}$) (this kinematic distance constraint relied on a value of 10\,kpc for the orbital radius of the sun with respect to the galactic centre).

More recent high-resolution SGPS data allowed TL2012 to not only find partial HI absorption at \vlsr$\sim -$65\,km\,s$^{-1}$, but also a distinct lack of HI absorption towards previously-overlooked gas at \vlsr$\sim -$145\,km\,s$^{-1}$, as traced by CO(1-0) \citep{Reynoso:2000}. TL2012 suggest that since the \vlsr$\sim -$65\,km\,s$^{-1}$ gas lies between the \vlsr$\sim -$110\,km\,s$^{-1}$ (HI absorption) and \vlsr$\sim -$145\,km\,s$^{-1}$ (no HI absorption) gas, the \vlsr$\sim -$65\,km\,s$^{-1}$ gas may plausibly be in a non-circular orbit, as would be the case if CTB\,37A is within the inner 3\,kpc of the galaxy, influenced by the gravitational potential of the Perseus arm. This implies that the CTB\,37A SNR distance may be between 6.3 and 9.5\,kpc, thus we assume a distance of 7.9$\pm$1.6 in our analyses. HI absorption information is summarised in Table\,\ref{Table:Absorption}.
\begin{table}
\centering
\small
\caption{A summary of previous HI absorption analyses towards CTB\,37A. \label{Table:Absorption}} 
\begin{tabular}{|l|c|l|}
	\hline
Gas Velocity& Absorption? & Relation to CTB\,37A\\
(km\,s$^{-1}$)& &\\
  \hline
$-$22	& $\checkmark$ 	& Foreground$^{1,2}$\\
$-$65	& partial		& Background/Some Foreground$^{1,2}$\\
$-$85	& $\checkmark$ 	& Foreground$^{1,2}$\\
$-$110	& $\checkmark$ 	& Foreground$^{1,2}$\\
$-$145	& $\times$ 		& Background$^2$\\
	\hline
\end{tabular}
\textit{$^{1}$\citet{Caswell:1975}},
\textit{$^{2}$\citet{Tian:2012}}
\end{table}

\section{Observations}\label{sec:Obs}
In April of 2010, we observed and co-added 5 Mopra OTF (on the fly) 19$^{\prime}\times$19$^{\prime}$ area maps at 7\,mm wavelength, centered on [$l,b$]=[348.43, 0.16], to produce a data cube with 2 spatial and 1 spectral (velocity) dimension. The scan length was 15.6$^{\prime\prime}$ per cycle time (of 2.0\,s) and the spacing between scan rows was 26$^{\prime\prime}$. Deep ON-OFF switched pointings were also taken towards six locations (see Figure\,\ref{fig:CTB1} and Table\,\ref{Table:DeepPointings}) containing 1720\,MHz OH masers and one location centered on a peak of CS(1-0) emission seen in mapping data (Location\,7). The naming convention for Locations\,1-6 corresponds to the order of presentation of OH masers chosen from Table\,2 of \citet{Frail:1996}.  One 12\,mm deep ON-OFF switched pointing was also performed towards the location [$l$,$b$]=[348.37$^{\circ}$,0.14$^{\circ}$] to follow up a feature discussed in \S\,\ref{sec:infra}. See \citet{Walsh:2008} for the 12\,mm spectrometer setup. We used a sky reference position of [$l,b$]=[346.40, $-$1.64].

The Mopra spectrometer, MOPS, was employed and is capable of recording sixteen tunable, 4096-channel (137.5\,MHz) bands simultaneously when in `zoom' mode, as used here. A list of measured frequency bands, targeted molecular transitions and achieved $\Trms$ levels are shown in Tables \ref{Table:bands} and \ref{Table:DeepPointings}.

\begin{table}
\centering
\small
\caption{The window set-up for the Mopra Spectrometer (MOPS) at 7\,mm. The centre frequency, targeted molecular line, targeted frequency and total efficiency-corrected map noise (\Trms) are displayed. \label{Table:bands}} 
\begin{tabular}{|l|l|l|l|l|}
	\hline
Centre& Molecular & \Trms & Detected?\\
Frequency&Emission Line& (K\,ch$^{-1}$) &(Map/Point)\\
(GHz)& & & \\
	\hline
42.310 				& $^{30}$SiO(J=1-0,v=0) 		& 0.07 & $\times$ $\times$\\
42.500				& $^{28}$SiO(J=1-0,v=3)			& 0.07 & $\times$ $\times$	\\
42.840 				& $^{28}$SiO(J=1-0,v=2) 		& 0.07 & $\times$ $\times$	\\
~~~$^{\prime \prime}$	& $^{29}$SiO(J=1-0,v=0) 		& ~~$^{\prime \prime}$ & $\times$ $\times$	\\
43.125 				& $^{28}$SiO(J=1-0,v=1) 		& 0.07 & $\times$ $\times$	\\
43.255 				& ~~~~~~- 							& 0.07 & 	\\
43.395 				& $^{28}$SiO(J=1-0,v=0) 		& 0.07 & $\times$ $\checkmark$	\\
44.085 				& CH$_3$OH(7(0)-6(1)\,A++) 		& 0.08 & $\times$ $\checkmark$	\\
44.535 				& ~~~~~~- 							& 0.08 & 	\\
45.125 				& HC$_7$N(J=40-39) 				& 0.09 & $\times$ $\times$ 	\\
45.255 				& HC$_5$N(J=17-16) 				& 0.09 & $\times$ $\times$ 	\\
45.465 				& HC$_3$N(J=5-4,F=5-4) 			& 0.09 & $\checkmark$ $\checkmark$	\\
46.225 				& $^{13}$CS(1-0) 				& 0.09 & $\times$ $\times$ 	\\
47.945 				& HC$_5$N(J=16-15) 				& 0.12 & $\times$ $\times$ 	\\
48.225 				& C$^{34}$S(1-0) 				& 0.12 & $\times$ $\times$ 	\\
48.635 				& OCS(J=4-3) 					& 0.13 & $\times$ $\times$ 	\\
48.975 				& CS(1-0) 					& 0.13 & $\checkmark$ $\checkmark$ 	\\
   \hline
\end{tabular}
\end{table}


The velocity resolution of the 7\,mm zoom-mode data is $\sim$0.2\,km\,s$^{-1}$. The beam FWHM and the pointing accuracy of Mopra at 7\,mm are 59$\pm$2$^{\prime\prime}$ and $\sim$6$^{\prime\prime}$, respectively. The achieved $\Trms$ values for the total 7\,mm map and individual pointings are stated in Tables \ref{Table:bands} and \ref{Table:DeepPointings}, respectively.

OTF-mapping and deep ON-OFF switched pointing data were reduced and analysed using the ATNF analysis programs, \textsc{Livedata}, \textsc{Gridzilla}, \textsc{Kvis}, \textsc{Miriad} and \textsc{ASAP}\footnote{See http://www.atnf.csiro.au/computing/software/}.

\textsc{Livedata} was used to calibrate data against a sky reference position measured after each scan of a row/column was completed. A polynomial baseline-subtraction was also applied. \textsc{Gridzilla} then combined corresponding frequency bands of multiple OTF-mapping runs into 16 three-dimensional data cubes, converting frequencies into line-of-sight velocities. Data were then combined, weighting by the Mopra system temperature and smoothed in the Galactic $l-b$ plane using a Gaussian of FWHM 1.25$^{\prime}$.

\textsc{Miriad} was used to correct for the efficiency of the instrument \citep{Urquhart:2010} for mapping data and create line-of-sight velocity-integrated intensity images (moment 0) from data cubes.

\textsc{ASAP} was used to analyse deep ON-OFF switched pointing data. Data were time-averaged, weighted by system temperature and had fitted polynomial baselines subtracted. Like mapping data, deep pointing spectra were corrected for the Mopra beam efficiency \citep{Urquhart:2010}. The beam efficiency for the CS(1-0) band for point and extended sources are $\sim$0.43 and $\sim$0.56, respectively.

CO(2-1) data was taken with the Nanten2 4\,m submillimeter telescope during December of 2008. The telescope has a beam FWHM of $\sim$90$^{\prime\prime}$ at 230\,GHz and a pointing accuracy of $\sim$15$^{\prime\prime}$. The Acoustic Optical Spectrometer (AOS) had 2048 channels separated by 0.38\,km\,s$^{-1}$, providing a bandwidth of 390\,km\,s$^{-1}$. The achieved $\Trms$ was $\lesssim$0.7\,K\,ch$^{-1}$. This CO(2-1) data offers superior spatial and velocity resolution when compared to previously-published CO(1-0) \citep{Dame:2001} data.

\section{Gas Parameter Calculation}\label{sec:GasParam}
To address the amount of available hadronic target material towards HESS\,J1714$-$385, the column density towards absorbed X-ray sources and the density towards the observed OH masers, we calculated column density, mass and density using CO(2-1), CS(1-0) and HI data. As we discuss in \S\,\ref{sec:GasDist}, the gas observed towards CTB\,37A appears extended, so we integrate our spectral line maps over a spatially wide region ([$\alpha$,$\delta$]-space) to estimate average gas parameters within chosen regions. We used the \textsc{Miriad} functions \textit{moment 0} and \textit{mbspect} to calculate integrated intensity, $\int_V\,T_b\,d\vlsr\,d\alpha\,d\delta$ (K\,km\,s$^{-1}$deg$^2$), and produce spectral line maps. 

The mass of gas within the region, $M$, is related to the average column density, \Nh, by 
\begin{equation}\label{equ:Mass}
M=2 m_H \Nh A
\end{equation}where $A$ is the cross-sectional area of the region and $2 m_H$ is the mass of molecular hydrogen. The average density, \nh, was estimated by assuming that the thickness of the region in the line of sight direction was of the same order as the height and width, $\sim\sqrt{A}$.

\subsection{CO}
We scaled our CO(2-1) emission by the expected CO(1-0)/CO(2-1) intensity ratio,
\begin{equation}
\label{equ:X21}
\frac{W_{1-0}}{W_{2-1}} 
\approx \frac{1}{4} \exp{\left[\frac{11.07\,\textrm{K}}{\Trot}\right]}
\end{equation}where $\Trot$ is the rotational temperature, which we assume to be 10\,K. This allowed us to use the commonly cited/measured CO(1-0) X-factor, $X_{1-0}\sim$3$\times$10$^{20}$\,cm$^{-2}$(K\,km\,s$^{-1}$)$^{-1}$ \citep{Dame:2001} to calculate H$_2$ column density, $N_{H_2}=X_{1-0}W_{1-0}$ from CO(2-1) intensity, $W_{2-1}$. Note, that a 0.7-3$\times$ systematic error in column density would be introduced for a 5\,K error in temperature estimation (using Equation\,\ref{equ:X21}). In an extreme case, such as $T=$40\,K, like towards the shock of similar SNR, SNR W28 \citep{Nicholas:2011}, we would expect the column density calculated from CO(2-1) using our method to change by a factor $\sim$0.4.


\subsection{CS}
We calculate the CS(J=1) column density using Equation\,9 from \citet{Goldsmith:1999}. This equation requires an optical depth, which would usually be estimated from the CS(1-0)-C$^{34}$S(1-0) intensity ratio, but as we do not detect C$^{34}$S(1-0), for simplicity we assume that the CS(1-0) emission is optically thin ($\tau\rightarrow$0). It follows that the column densities calculated from CS(1-0) may be considered as lower limits.

An LTE assumption at \Trot$\sim$10\,K implies that the total CS column density is simply a factor $\sim$3.5 times the CS(J=1) column density, and the molecular abundance of CS with respect to hydrogen is assumed to be $\sim$10$^{-9}$ \citep{Frerking:1980}. The systematic error introduced into the CS(1-0) column density by our temperature assumption is likely to be even smaller than that introduced for CO(2-1), with a 50\% temperature systematic corresponding to a factor 0.7-1.2 error in column density. In the extreme case of $T=$40\,K, column density calculated from CS(1-0) using our method changes by $\sim$1.8 times.



\subsection{HI}
HI data is available \citep{McClure:2005,Haverkorn:2006}, and exhibits emission and absorption features towards CTB\,37A. Where the level of radio continuum-absorption appears minor and HI emission is present, we constrain column density using an HI X-factor, $X_{HI}=$1.823$\times$10$^{18}$\,cm$^{-2}$(K\,km\,s$^{-1}$)$^{-1}$ \citep{Dickey:1990}.

For absorption lines, we first estimate an optical depth, $\tau\sim\ln\left(T_0 / T_b \right) $, where $T_0$ is the continuum intensity and $T_b$ is the local minimum intensity of an absorption line. This allows the calculation of atomic column density, $N_H\sim$1.9$\times$10$^{18}\tau\fwhm T_s$, where $\fwhm$ is an approximated full-width-half-maximum and $T_s$ is the spin temperature, assumed to be 100\,K. Since the calculated column density is proportional to the spin temperature, the uncertainty in column density due to our assumed temperature is the percentage as the uncertainty in temperature.



\section{The Spatial Gas Distribution}\label{sec:GasDist}
Figure\,\ref{fig:PVplot} is a longitude-velocity plot of CO(2-1) and CS(1-0) data (see Figure\,\ref{fig:PVplot_lat} for a latitude-velocity plot). Several clouds are visible in CO(2-1) and CS(1-0) emission at approximate line-of-sight reference velocities, $\vlsr\sim$ 5, $-$10, $-$22, $-$60 to $-$75, $-$90 and $-$105\,km\,s$^{-1}$. Additionally, SiO(1-0), HC$_3$N(J=5-4) and CH$_3$OH(7(0)-6(1)\,A$++$) were detected towards `Location\,3' (Figure\,\ref{fig:Spec3}) at velocity, $\vlsr\sim -$105\,km\,s$^{-1}$. Gaussian line fit parameters to the most significant emission lines are presented in Table\,\ref{Table:DeepPointings}.

\begin{figure}
\centering
\includegraphics[width=0.48\textwidth]{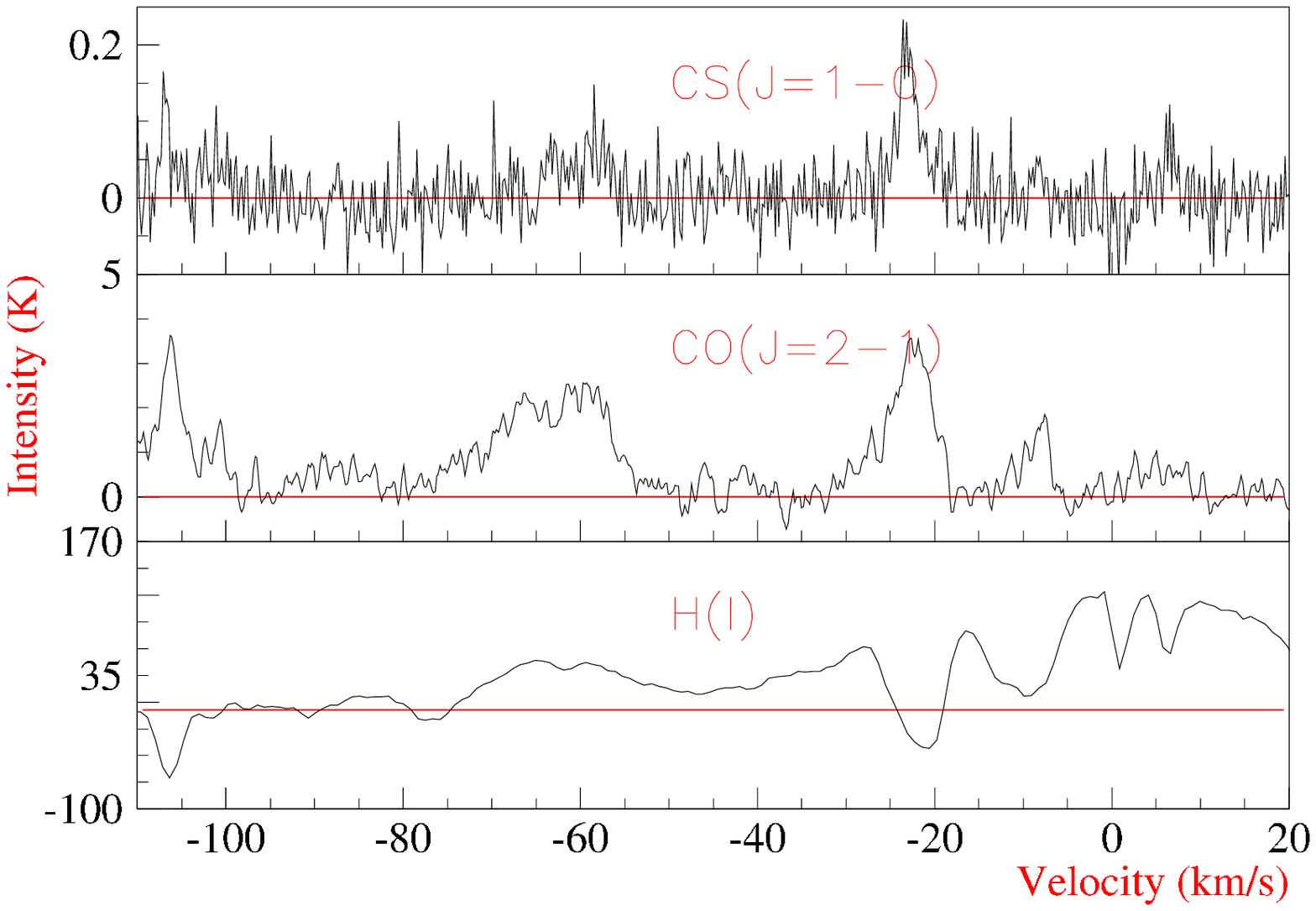}
\caption{\textbf{Location\,5:} This region encompasses two 1720\,MHz OH masers. One is at velocity \vlsr$\sim$\,$-$23.3\,km\,s$^{-1}$ (the centre of the pointing location) and another velocity \vlsr$\sim$\,$-$21.4\,km\,s$^{-1}$ (offset from the centre, but within the 7\,mm beam FWHM solid angle). See Table\,\ref{Table:DeepPointings} and Figure\,\ref{fig:CTB1}b for further details. \label{fig:Spec5}}
\end{figure}

\begin{figure}
\centering
\includegraphics[width=0.48\textwidth]{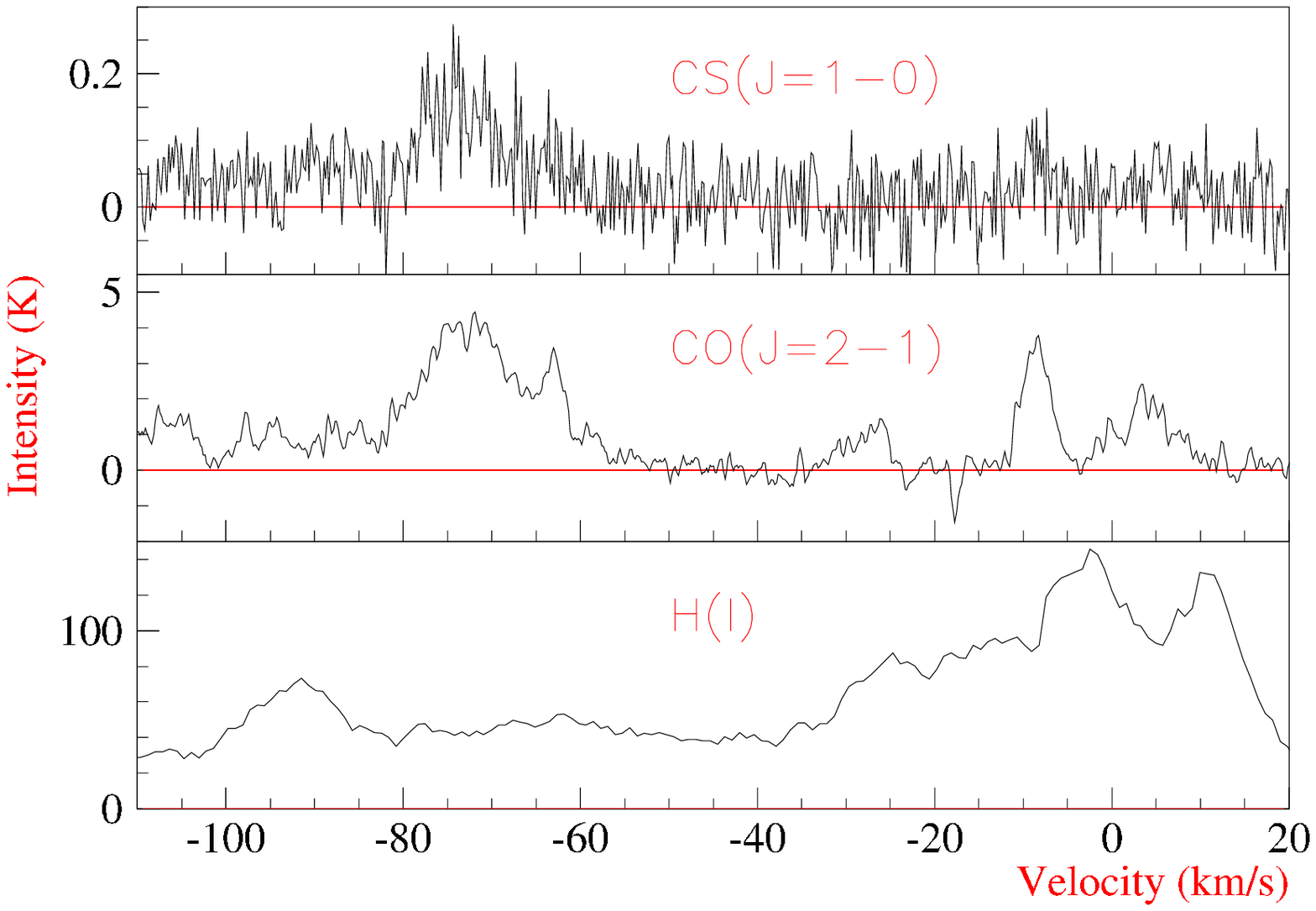}
\caption{\textbf{Location\,1:} This region encompasses one 1720\,MHz OH maser at velocity \vlsr$\sim$\,$-$65.1\,km\,s$^{-1}$. See Table\,\ref{Table:DeepPointings} and Figure\,\ref{fig:CTB1}b for further details.\label{fig:Spec1}}
\end{figure}

\begin{figure}
\centering
\includegraphics[width=0.48\textwidth]{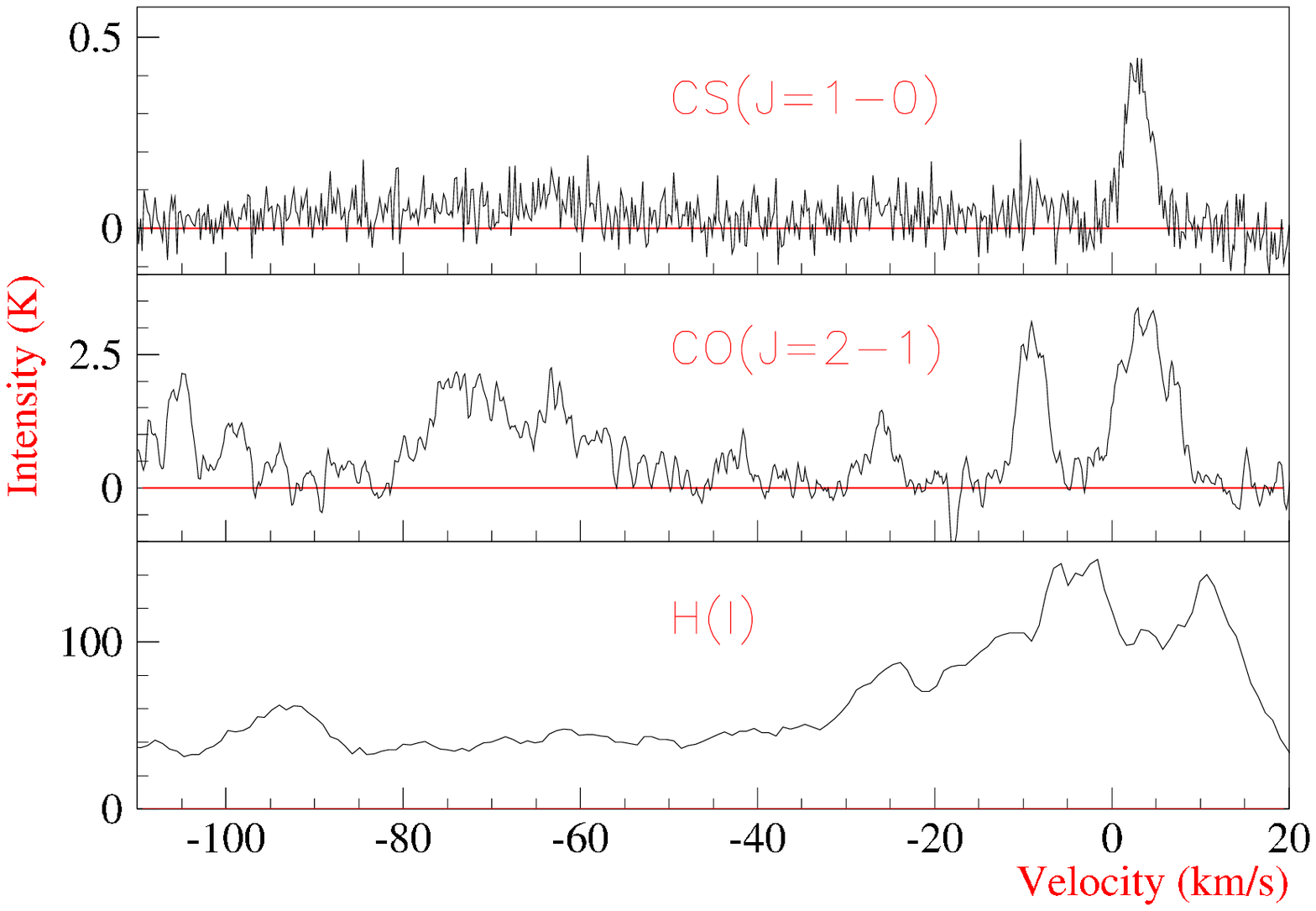}
\caption{\textbf{Location\,2:} This region encompasses one 1720\,MHz OH maser at velocity \vlsr$\sim$\,$-$65.2\,km\,s$^{-1}$. See Table\,\ref{Table:DeepPointings} and Figure\,\ref{fig:CTB1}b for further details.\label{fig:Spec2}}
\end{figure}

\begin{figure}
\centering
\includegraphics[width=0.48\textwidth]{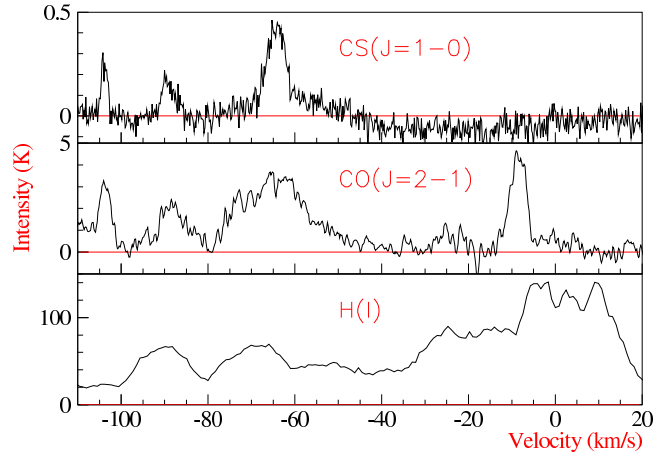}
\caption{\textbf{Location\,7:} See Table\,\ref{Table:DeepPointings} and Figure\,\ref{fig:CTB1}b for further details.\label{fig:SpecCS}}
\end{figure}

Together, Mopra CS(1-0), Nanten CO(2-1) and SGPS HI data \citep{McClure:2005,Haverkorn:2006} towards seven CTB\,37A locations are shown in Figures\,\ref{fig:Spec5}-\ref{fig:SpecCS}. The HI data contained a strong radio continuum component from CTB\,37A, producing absorption lines in foreground gas.

\begin{table*}
\caption{Detected molecular transitions from deep ON-OFF switched pointings. Velocity of peak, \vlsr, peak intensity, \Tpeak, and FWHM, \fwhm, were found by fitting Gaussian functions before deconvolving with MOPS velocity resolution. Displayed band noise, \Trms, and peak temperatures take into account beam efficiencies and contain a $\sim$5\% systematic uncertainty \protect\citep{Urquhart:2010}, after a baseline subtraction. \label{Table:DeepPointings}} 
\begin{tabular}{|l|l|c|r|c|r|l|l|}
	\hline
Object			&Detected	   	& \Trms		&Peak \vlsr		& \Tpeak		&\fwhm 		&Counterpart\\
(l,b)			&Emission Line	&(K\,ch$^{-1}$)		&(km\,s$^{-1}$)	& (K)           &(km\,s$^{-1}$)&	\\
\hline
Location 1		& CS(1-0)		& 0.05		&$-$73.4$\pm$0.5	&0.13$\pm$0.01 	&18.9$\pm$1.7&$-$65.1\,km\,s$^{-1}$ OH maser\\
(348.35,0.18)&				&			&				&				& 			& \\
\hline
Location 2		& CS(1-0)		& 0.06		&2.87$\pm$0.08	&0.41$\pm$0.02	&3.6$\pm$0.2 &-65.2\,km\,s$^{-1}$ OH maser\\
(348.35,0.15)& HC$_3$N(J=5-4,F=5-4)&0.05	&4.7$\pm$0.2	&0.12$\pm$0.03	&1.9$\pm$0.6 &\\
\hline
Location 3		& CS(1-0)		& 0.05		&$-$105.1$\pm$0.03&0.78$\pm$0.02	&2.5$\pm$0.1 &$-$63.8\,km\,s$^{-1}$ OH maser$^{a}$\\
(348.42,0.11)&				&			&$-$88.0$\pm$0.01	&0.21$\pm$0.02	&1.6$\pm$0.2 &\\
				&				&			&$-$57.6$\pm$0.4	&0.17$\pm$0.01	&21.8$\pm$1.1&	\\
				&				&			&$-$22.4$\pm$0.5	&0.10$\pm$0.07	&9.2$\pm$1.4 &	\\
				& SiO(J=1-0,v=0)& 0.03		&$-$104.6$\pm$0.3	&0.06$\pm$0.01	&4.4$\pm$1.2 &	\\
				& 				& 			&$-$60.2$\pm$2.0	&0.02$\pm$0.05	&19.2$\pm$5.4 &	\\				
			& HC$_3$N(J=5-4,F=5-4)& 0.03		&$-$104.8$\pm$0.1	&0.20$\pm$0.01	&2.7$\pm$0.3 &	\\
& CH$_3$OH(7(0,7)-6(1,6)\,A++)	& 0.03		&$-$104.4$\pm$0.1	&0.13$\pm$0.01	&1.9$\pm$0.2 &	\\				
\hline
Location 4		& CS(1-0)		& 0.04		&$-$61.9$\pm$0.6	&0.07$\pm$0.01	&12.1$\pm$1.8 &$-$63.5\,km\,s$^{-1}$ OH maser\\
(348.39,0.08)&				& 			&$-$21.9$\pm$0.2	&0.13$\pm$0.02	&1.5$\pm$0.3 &\\
				&				& 			&8.3$\pm$0.1	&0.14$\pm$0.02	&1.5$\pm$0.3 &\\
\hline
Location 5		& CS(1-0)		& 0.04		&$-$23.0$\pm$0.1	&0.18$\pm$0.02	&3.1$\pm$0.4 &$-$23.3\,km\,s$^{-1}$ OH maser$^{a}$\\
(348.48,0.11)&				&			&				&				& &\\
\hline
Location 6		& CS(1-0)		& 0.07		&$-$63.9$\pm$0.2	&0.23$\pm$0.02	&7.4$\pm$0.6 &$-$64.3\,km\,s$^{-1}$ OH maser\\
(348.51,0.11)&				&			&				&				& &\\
\hline
Location 7		& CS(1-0)		& 0.05		&$-$103.9$\pm$0.1	&0.26$\pm$0.02	&1.7$\pm$0.2 &		\\		
(348.32,0.23)	&				&			&$-$89.2$\pm$0.2	&0.16$\pm$0.02	&3.7$\pm$0.4 &	\\
				&				&			&$-$64.2$\pm$0.1	&0.41$\pm$0.01	&6.1$\pm$0.3 &	\\
\hline
\end{tabular}
$^{a}$Pointing has partial overlap with a second 1720\,MHz OH maser of similar \vlsr.
\end{table*}

\subsection{Gas Morphology}
\begin{figure}
\centering
\includegraphics[width=0.45\textwidth]{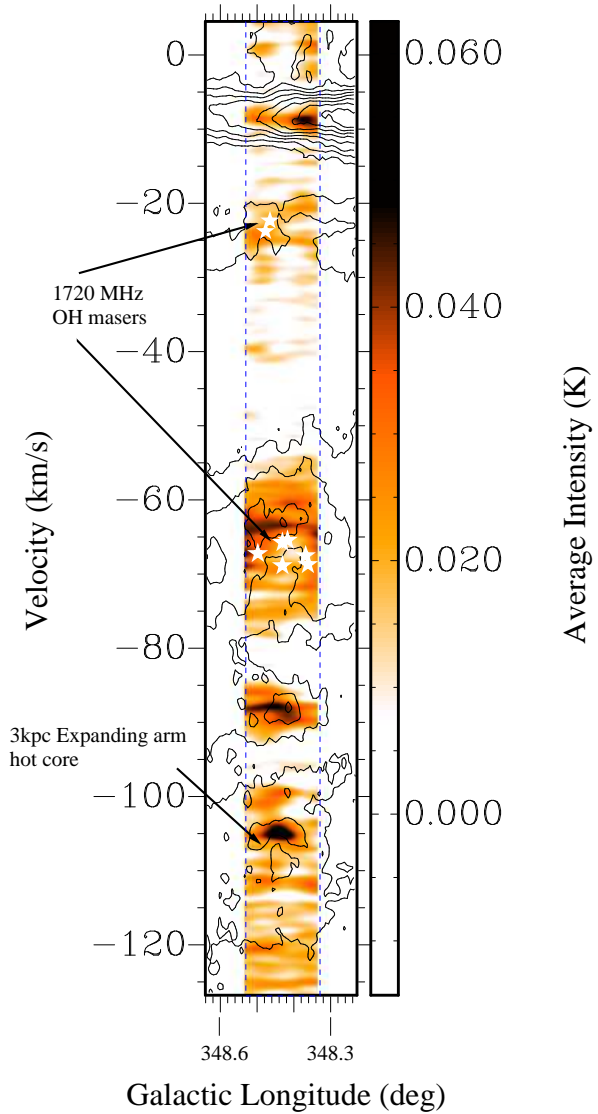}
\caption{Position-velocity image of Mopra CS(1-0) emission towards the CTB\,37A region. Nanten CO(2-1) emission contours towards the CTB\,37A region are overlaid. 
White stars indicate 1720\,MHz OH maser locations and the dashed lines indicate the extent of the 7\,mm mapping campaign for CS(1-0) emission. A position-velocity image for the latitudinal gas distribution is displayed in the appendix (\S\ref{fig:PVplot_lat}). The image has been smoothed for clarity.\label{fig:PVplot}}
\end{figure}

\begin{figure*}
\centering
\includegraphics[width=1.02\textwidth]{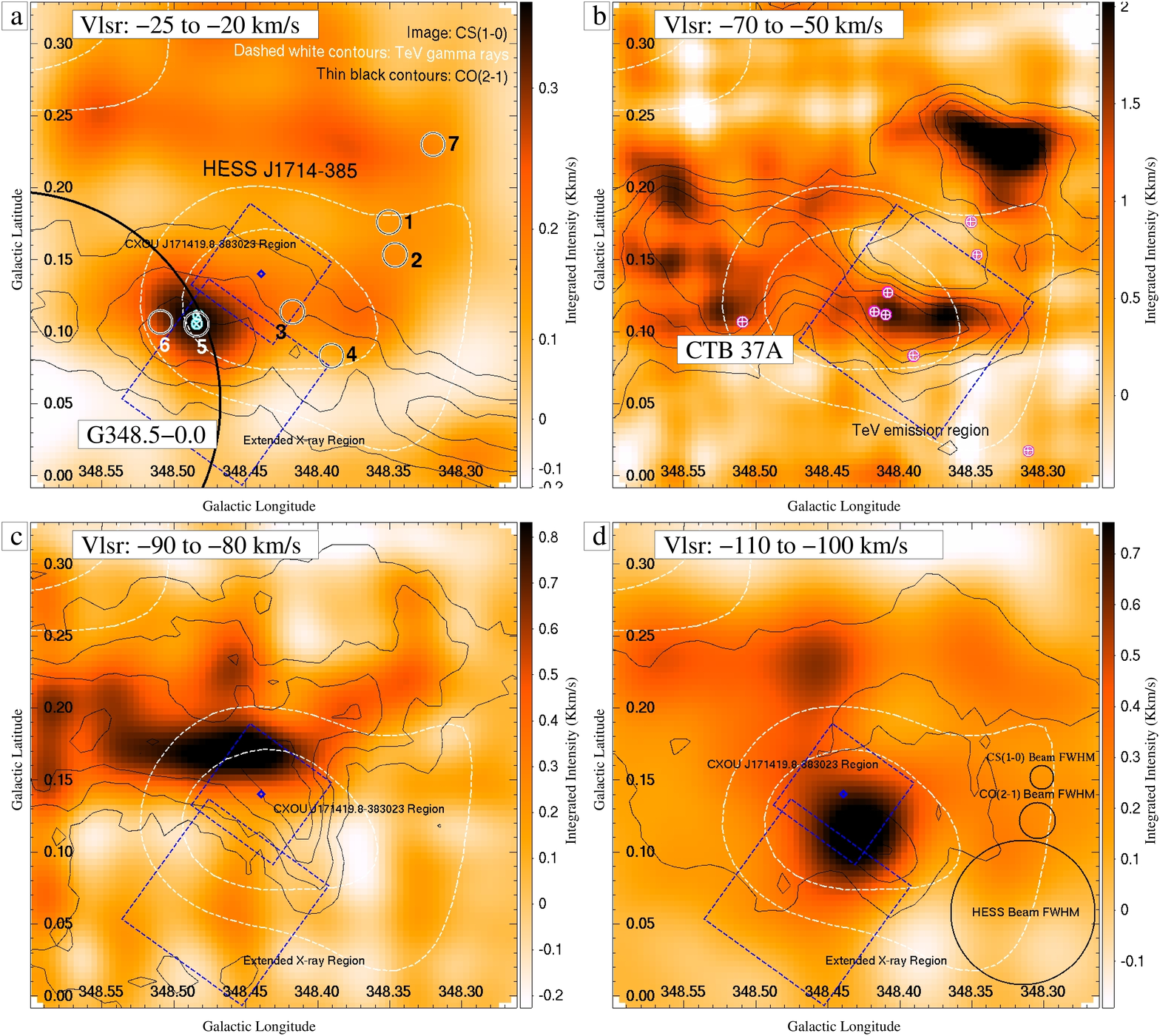}
\caption{Mopra CS(1-0) emission images with overlaid Nanten2 CO(2-1) contours (black) for four different velocity ranges (panels a-d). CS(1-0) emission in all figures has been Gaussian-smoothed (FWHM$\sim$30-90$^{\prime\prime}$). All figures have overlaid HESS 80 and 100 gamma-ray excess count contours (dashed white lines) \protect\citep{Aharonian:2008}. \textbf{(a)} CO contour levels are from 5 to 20\,K\,km\,s$^{-1}$ in steps of 5\,K\,km\,s$^{-1}$. The approximate location of G348.5$-$0.0 is indicated by a black circle (partially shown). Two blue dashed boxes represent regions used to calculate gas parameters (see \S\,\ref{sec:X-ray}, also in c and d). Deep switched pointing locations are displayed as circles and 1720\,MHz OH masers at the velocity of the CS(1-0) emission are indicated by blue crosses. \textbf{(b)} CO contour levels are from 24 to 48\,K\,km\,s$^{-1}$ in steps of 8\,K\,km\,s$^{-1}$. A large blue dashed box (TeV emission) represent a region used to calculate gas parameters (see \S\ref{sec:tev}). 1720\,MHz OH masers at the velocity of the CS(1-0) emission are indicated by pink crosses. \textbf{(c)} CO contour levels are from 5 to 20\,K\,km\,s$^{-1}$ in steps of 5\,K\,km\,s$^{-1}$. \textbf{(d)} CO contour levels are from 10 to 30\,K\,km\,s$^{-1}$ in steps of 10\,K\,km\,s$^{-1}$. The beam FWHM for HESS, Mopra and Nanten2 observations are indicated. \label{fig:CSpics}}
\end{figure*}

Referring to \citet{Moriguchi:2005}, which addresses CO(1-0) emission from the nearby SNR RX\,J1713.7$-$3946, regions traced by CO(2-1) and CS(1-0) may correspond to local gas ($\sim$5\,km\,s$^{-1}$), the Sagittarius arm ($\sim-$5 to $-$30\,km\,s$^{-1}$), the Norma arm ($\sim$\,$-$60 to $-$75\,km\,s$^{-1}$), `Cloud\,A' ($\sim$\,$-$80 to $-$95\,km\,s$^{-1}$, \citeauthor{Slane:1999}, \citeyear{Slane:1999}) and the 3\,kpc expanding arm ($\sim$\,$-$100 to $-$115\,km\,s$^{-1}$). Modeling by \citet{Vallee:2008} is not necessarily consistent with this scenario, with Figure\,3d illustrating a model with possible gas components of the Sagittarius-Carina, Scutum-Crux and Norma-3\,kpc arms present at velocities \vlsr$\sim$\,$-$30 to 0\,km\,s$^{-1}$ towards CTB\,37A (\citeauthor{Vallee:2008} also unified the various naming conventions for different components of single arms, including the Norma and 3\,kpc arms which were part of the same structure). For velocities less than \vlsr$\sim$\,$-$40\,km\,s$^{-1}$, the model may predict the presence of components of the Perseus and Norma-3\,kpc arms, but, as discussed by the authors, the model is less reliable towards the central 22$^{\circ}$ of the Galactic plane, so we do not attempt to rectify it with our observations.

In contrast to the above discussion, TL2012 make a case for the gas at $-$20\,km\,s$^{-1}$ and $-$60\,km\,s$^{-1}$ to both be associated with CTB\,37A (i.e. present in the same arm) and within the inner 3\,kpc of the galaxy (from HI-absorption arguments, see \S\,\ref{sec:Distance}).

All CO(2-1) emission in Figure\,\ref{fig:PVplot} appears to have corresponding CS(1-0) emission, indicating dense gas. 
CS(1-0) and CO(2-1) emission integrated around the two separate populations of OH masers at $\vlsr\sim-$22 and $-$65\,km\,s$^{-1}$ are displayed in Figures\,\ref{fig:CSpics}a and b. Figures\,\ref{fig:CSpics}c and d display two other regions of gas along the line of sight that appear in HI absorption. In Figure\,\ref{fig:CSpics}, CS(1-0) emission peaks in regions displaying intense CO(2-1) emission. We investigate the association of this gas with CTB\,37A and G348.5$-$0.0 in the following sections.

\subsection{G348.5$-$0.0 and Interstellar Gas at \vlsr =$-$25 to $-$20\,km\,s$^{-1}$}\label{sec:G348}
Figure\,\ref{fig:CSpics}a shows CS emission integrated over a velocity containing two 1720\,MHz OH masers at \vlsr$\sim$\,$-$23 and $\sim$\,$-$21\,km\,s$^{-1}$\citep{Frail:1996}. The CS(1-0) emission is coincident with these OH masers and a CO(2-1) emission peak and is somewhat extended, spanning $\sim$240$^{\prime\prime}$ ($\sim$4$\times$Beam FWHM) in length. Also note that the CS(1-0) emission peak may lie on the edge of G348.5$-$0.0, as approximately extrapolated from the partial shell observed in 843\,MHz continuum data (see Figure\,\ref{fig:CTB1}). This is consistent with the \vlsr$\sim$\,$-$22\,km\,s$^{-1}$ masers being a product of a G348.5$-$0.0 shock-interaction towards a region of high density.

\begin{figure}
\centering
\includegraphics[width=0.48\textwidth]{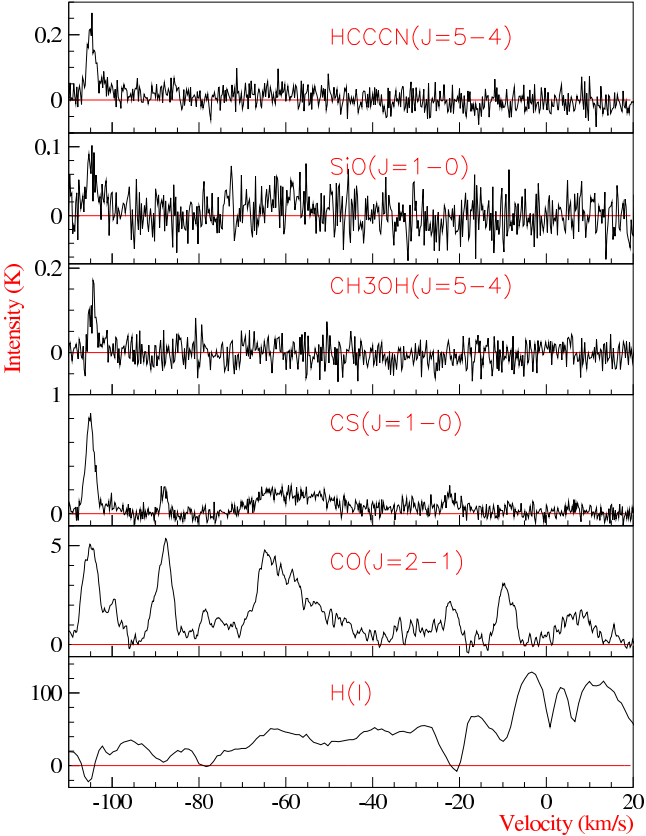}
\caption{\textbf{Location\,3:} This region encompasses two 1720\,MHz OH masers. One is at velocity \vlsr$\sim$\,$-$63.8\,km\,s$^{-1}$ (the centre of the pointing location) and another velocity \vlsr$\sim$\,$-$63.9\,km\,s$^{-1}$ (offset from the centre, but within the 7\,mm beam FWHM solid angle). See Table\,\ref{Table:DeepPointings} and Figure\,\ref{fig:CTB1}b for further details.\label{fig:Spec3}}
\end{figure}

\begin{figure}
\centering
\includegraphics[width=0.48\textwidth]{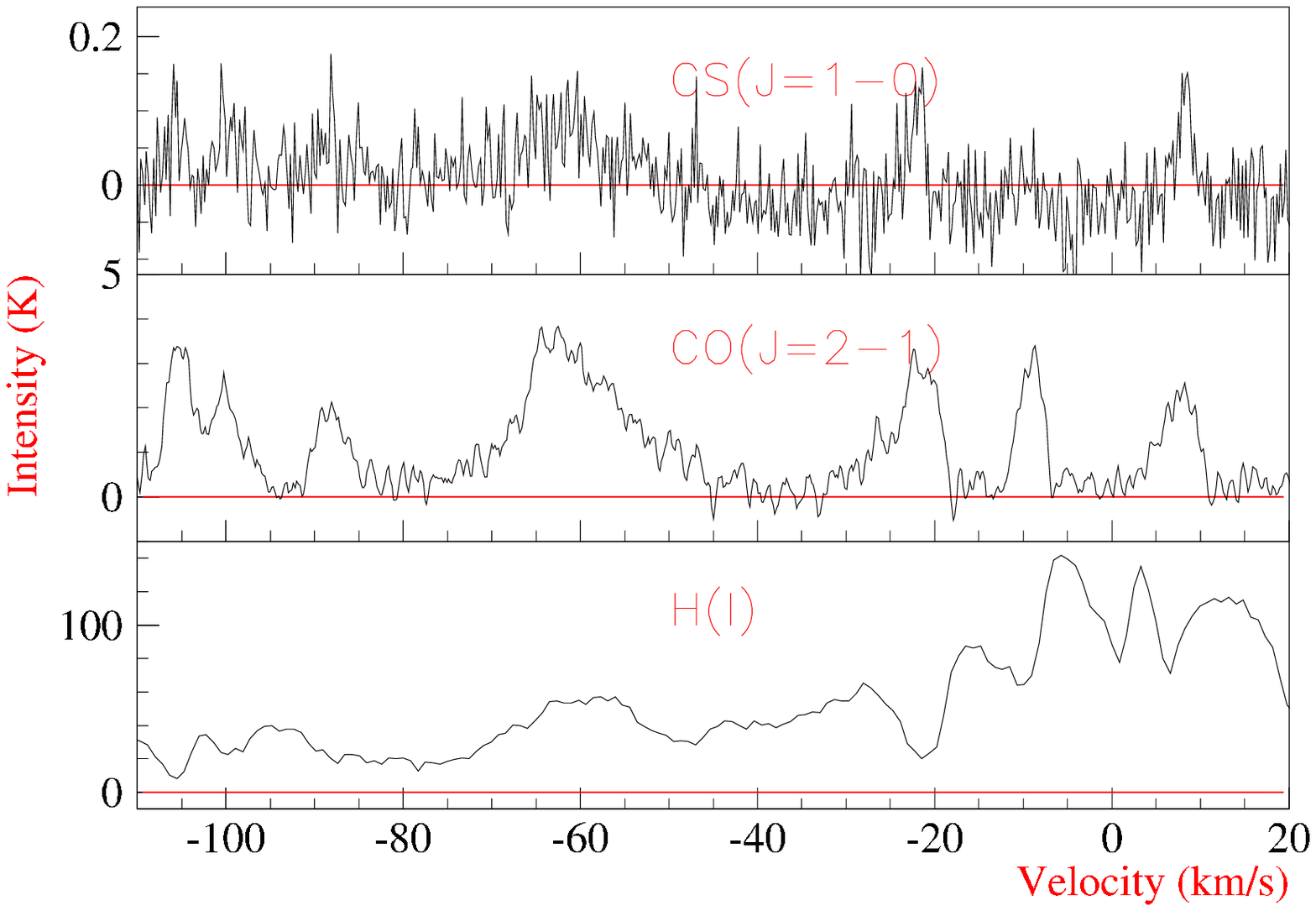}
\caption{\textbf{Location\,4:} This region encompasses one 1720\,MHz OH maser at velocity \vlsr$\sim$\,$-$63.5\,km\,s$^{-1}$. See Table\,\ref{Table:DeepPointings} and Figure\,\ref{fig:CTB1}b for further details. \label{fig:Spec4}}
\end{figure}

\begin{figure}
\centering
\includegraphics[width=0.48\textwidth]{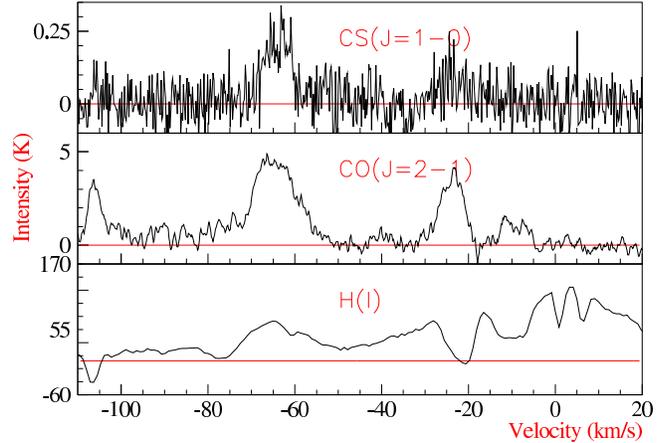}
\caption{\textbf{Location\,6:} This region encompasses one 1720\,MHz OH maser at velocity \vlsr$\sim$\,$-$64.3\,km\,s$^{-1}$. See Table\,\ref{Table:DeepPointings} and Figure\,\ref{fig:CTB1}b for further details.\label{fig:Spec6}}
\end{figure}

Figure \ref{fig:Spec5} displays CO(2-1), CS(1-0) and HI (and continuum) data towards `Location\,5', which contains the \vlsr$\sim$\,$-$22\,km\,s$^{-1}$ population of 1720\,MHz OH masers. CS(1-0) emission corresponds to at least three of the most intense CO(2-1) peaks (\vlsr$\sim$\,$-$22, $-$65, $-$105\,km\,s$^{-1}$). The strongest CS(1-0) emission peak towards this location is at velocity \vlsr$\sim$\,$-$23\,km\,s$^{-1}$, consistent with the velocity of the OH masers. A distinct HI absorption line also corresponds to the dense and masing gas at \vlsr$\sim$\,$-$22\,km\,s$^{-1}$. This implies that the gas is likely foreground to the radio continuum emission from this region. It follows that if the radio continuum emission seen towards Location\,5 is indeed from CTB\,37A, the \vlsr$\sim$\,$-$22\,km\,s$^{-1}$ gas (and associated OH maser population) lies in front of CTB\,37A. Similar \vlsr$\sim$\,$-$22\,km\,s$^{-1}$ HI absorption features are observed towards Locations 3, 4 and 6 (Figures \ref{fig:Spec3}, \ref{fig:Spec4} and \ref{fig:Spec6}, respectively). These positions are, unlike Location\,5, not towards regions of suspected SNR-SNR-overlap, only SNR CTB\,37A.

The \vlsr$\sim$\,$-$22\,km\,s$^{-1}$ gas may be shocked by either CTB\,37A or G348.5$-$0.0. In the former case, the gas would not be at its kinematic distance (possibly within the 3\,kpc ring instead), while CTB\,37A would be between the foreground \vlsr$\sim$\,$-$22\,km\,s$^{-1}$ gas and the background/associated \vlsr$\sim$\,$-$65\,km\,s$^{-1}$ gas.


In the latter case (a G348.5$-$0.0-\vlsr$\sim$\,$-$22\,km\,s$^{-1}$ maser/gas association), the Galactic rotation model distance may hold for the \vlsr$\sim$\,$-$22\,km\,s$^{-1}$ gas, favouring a near-side distance solution of $\sim$3\,kpc for the \vlsr$\sim$\,$-$22\,km\,s$^{-1}$ masers, molecular gas and G348.5$-$0.0. It is unclear from current evidence which of the two scenarios is valid, but 1720\,MHz OH maser emission from SNRs is relatively rare, with only $\sim$10\% of surveyed SNRs having detectable OH masers \citep{Green:1997}. Although the \citeauthor{Green:1997} survey did target (hence is biased towards) SNRs with a central thermal X-ray source, one might naively suggest, post-priori, that the likelihood of G348.5$-$0.0 producing OH maser emission is about 10\%, with the more likely cause of \vlsr$\sim$\,$-$22\,km\,s$^{-1}$ OH maser generation being an interaction between SNR CTB\,37A and a second molecular gas clump of a different local velocity (the \vlsr\,$\sim$22\,km\,s$^{-1}$ gas). The work by TL2012 does seem to support the notion that the kinematic distances of these gas components are not reliable.

An estimation of mass and density at \vlsr$\sim$\,$-$22\,km\,s$^{-1}$ towards Location\,5 can be found using Equation\,\ref{equ:Mass}, where $A$ is set equal to the Mopra beam FWHM area (at some assumed distance). For CS(1-0) emission, this yields a column density of $N_{H_2}\sim$1$\times$10$^{22}$\,cm$^{-2}$, which implies a mass of $\sim$400\,$M_{\odot}$ and density of $\sim$1$\times$10$^3$\,cm$^{-3}$ at a distance of 7.9\,kpc (or $\sim$40\,$M_{\odot}$ and $\sim$5$\times$10$^3$\,cm$^{-3}$ at a distance of 3\,kpc, if G348.5$-0.0$ produced the \vlsr$\sim$\,$-$22\,km\,s$^{-1}$ masers). This density is not as high as the $\sim$10$^4$-10$^5$\,cm$^{-3}$ suspected to be required for the formation of 1720\,MHz OH masers \citep{Reynoso:2000}, but our result does not rule out this density on a scale smaller than our 7\,mm beam ($\sim$2\,pc at 7.9\,kpc), as expected for the maser-emission region size. Indeed, given that the CS(1-0) critical density is $\sim$8$\times$10$^4$\,cm$^{-3}$, its likely that the beam is only partially-`filled'. Additionally, we have used an `optically-thin' assumption (see \S\ref{sec:GasParam}), but the region could indeed be more optically deep, hence have a higher column density than what we have calculated.

\subsection{CTB\,37A and Interstellar Gas at \vlsr =$-$70 to $-$50\,km\,s$^{-1}$}\label{sec:CTB37A}
Figure\,\ref{fig:CSpics}b shows CS(1-0) emission integrated over a velocity consistent with eight 1720\,MHz OH masers between \vlsr$\sim$\,$-$70 and $\sim$\,$-$60\,km\,s$^{-1}$ \citep{Frail:1996}. Broad CS(1-0) emission is seen towards regions with intense CO(1-0) emission. Some of the gas in this field is believed to be associated with the CTB\,37A SNR and the $\sim$\,$-$60\,km\,s$^{-1}$ masers, which are in the inner 3\,kpc of the Galaxy (between 6.3\,kpc and 9.5\,kpc, TL2012).  



The existence of a shocked or energetic environment at \vlsr$\sim -$65\,km\,s$^{-1}$, possibly triggered by CTB\,37A, is indicated by the extreme broadness of the CO(2-1) and CS(1-0) line spectra and clear asymmetries in CO(2-1) emission towards Locations 3 and 4 (Figures\,\ref{fig:Spec3} and \ref{fig:Spec4}). In such environments, the detection of SiO emission can sometimes be expected as Si is sputtered from dust grains, but the only signal corresponding to CTB\,37A from this shock-tracer was tenuous. A fit to the SiO(1-0) spectrum towards Location\,3 (Figure\,\ref{fig:Spec3}, Table\,\ref{Table:DeepPointings}) at \vlsr$\sim$\,$-$60\,km\,s$^{-1}$ (guided by the broad CS(1-0) emission) yielded a line FWHM of $\sim$19$\pm$5\,km\,s$^{-1}$, consistent with a shocked or turbulent region. This detection, albeit weak, may support the connection of the \vlsr$\sim -$65\,km\,s$^{-1}$ gas to the CTB\,37A SNR.

Via CS(1-0) emission we can confirm the existence of dense gas-associations with \vlsr$\sim$\,$-$65\,km\,s$^{-1}$ OH masers towards locations 3, 4 and 6. We estimate, column densities, masses and densities from CS(1-0) of $\sim$1-8$\times$10$^{22}$\,cm$^{-2}$, $\sim$500-2300\,$M_{\odot}$ and $\sim$3-10$\times$10$^{3}$\,cm$^{-3}$, respectively, at these three locations (with assumptions outlined in \S\,\ref{sec:GasParam}). These densities are, similar to the previous calculation (see \S\,\ref{sec:CTB37A}), not as high as the $\sim$10$^4$-10$^5$\,cm$^{-3}$ suspected to be required for the formation of 1720\,MHz OH masers. This does not rule out a density of $\sim$10$^4$-10$^5$\,cm$^{-3}$ on a scale smaller than our 7\,mm beam ($\sim$2\,pc at 7.9\,kpc).


Towards locations 1 and 2, no CS(1-0) emission was confirmed above the noise level of $\sim$0.05\,K, so like \citet{Reynoso:2000}, we are unable to give evidence of a dense-molecular gas association towards the OH masers at these two positions. Lastly, we note that towards these two positions, a shell-like structure of radius $\sim$0.025$^{\circ}$ centred near [$l$,$b$]$\sim$[348.37,$-$0.16] appears to exist in CS(1-0) and CO(2-1) emission. We are unable to determine the cause of this void in molecular emission, but this may be a low gas density cavity blown out by a SNR or SNR progenitor star wind. Furthermore, we note a distict lack of molecular gas to the South of CTB\,37A, so we are unable to add more to the discussion of the break-out morphology of CTB\,37A.

\subsection{Interstellar Gas at \vlsr$ =-$90 to $-$80\,km\,s$^{-1}$}\label{sec:-85}
Figure\,\ref{fig:CSpics}c is an integrated image of the CS(1-0) and CO(2-1) emission around \vlsr$ =-$85\,km\,s$^{-1}$. Some HI-absorption at \vlsr$\sim$\,$-$85\,km\,s$^{-1}$ is present towards Location\,3 (Figure\,\ref{fig:Spec3}). This is evidence that the \vlsr$\sim$\,$-$85\,km\,s$^{-1}$ gas is foreground to CTB\,37A, but this is not especially useful in distinguishing whether this gas is associated with the CTB\,37A SNR (including whether this cloud is experiencing a high energy particle enhancement caused by CTB\,37A). It follows that we consider both of two scenarios: an association of this gas with the gamma-ray emission and no association of this gas with the gamma-ray emission in \S\ref{sec:tev}. This gas appears at the same velocity as `Cloud\,A'  \citep{Slane:1999}, which can be seen to extend over 1$^{\circ}$ in longitude \citep{Moriguchi:2005}, suggesting a possible relation.

\subsection{Interstellar Gas at \vlsr =$-$110 to $-$100\,km\,s$^{-1}$}\label{sec:-110}
Gas from the near side of the 3\,kpc-expanding-arm is observed at \vlsr$\sim$\,$-$105\,km\,s$^{-1}$ in the direction of CTB\,37A, as displayed in Figure\,\ref{fig:CSpics}d. 
The significance of this gas to the present study is a correspondence with HI absorption in the CTB\,37A radio continuum (see \S\,\ref{sec:Distance}). 
CS(1-0) and CO(2-1) emission peak at a position overlapping Location\,3 at $\vlsr\sim-$105\,km\,s$^{-1}$. CS(1-0) and HC$_3$N(J=5-4) emission indicates that the gas is dense (CS(1-0) emission $\Rightarrow$ \nH$\sim$7$\times$10$^3$\,cm$^{-3}$, see Table\,\ref{Table:GasParamsPointings}), while CH$_3$OH(7(0)-6(1)\,A++) emission suggests the presence of high-temperature chemistry often associated with star-formation \citep{vanDishoeck:1998}. SiO(1-0) emission is also associated with this clump, implying the existence of shocked gas. Although this might be considered evidence for a SNR shock-gas interaction, the coinciding tracers (CS, HC$_3$N and CH$_3$OH) can also frame a picture consistent with in/outflows associated with star-formation (see Figure\,\ref{fig:CTB_Species} for SiO, HC$_3$N and CH$_3$OH emission maps). Indeed, this narrow SiO(1-0) spectrum does not share similarities with shocked gas towards the similar SNR W28, where broad SiO emission is observed at a position/velocity consistent with OH maser emission \citep{Nicholas7mm:2012}. We display the discovered SiO, HC$_3$N and CH$_3$OH emission and investigate possible infrared counterparts in appendices (\S\ref{sec:infra} and \S \ref{sec:addimages}).

\begin{landscape}
\begin{table*}
\centering 
\caption{Gas parameters towards a rectangular region of sky containing the HESS\,J1714-385 68\% containment region defined as ``TeV emission'' (see Figure\,\ref{fig:CSpics}b), extended X-ray emission and a PWN candidate source, CXOU\,J171419.8$-$383023, (see Figures\,\ref{fig:CSpics}a, c and d) \citep{Aharonian:2008}. CO(2-1) and CS(1-0)-derived average column density, $N_{H_2}$, mass, $M$, and average density, $n_{H_2}$, within the selected region are shown for chosen velocity ranges. All calculated values have an associated statistical uncertainty of $\sim$10\%. \label{Table:GasParams}}
\begin{tabular}{l|l|l|r|c|l|r|c|l}
\hline
Region	&	Velocity	&	 Assumed	&	 \multicolumn{3}{c}{CO(2-1)} 																				&	 \multicolumn{3}{c}{CS(1-0)}					\\
	&	Range	&	 Distance, $d_0$	&	 $\Nh$	&	  $M$	&	$\nh$ 	&															 $\Nh$ 	&	$M$	&	 \nh 	\\
	&	(km\,s$^{-1}$)	&	(kpc)$^b$	&	 (10$^{21}$\,cm$^{-2}$)	&	 (M$_{\odot}$)	&	 (cm$^{-3}$) 	&															 (10$^{21}$\,cm$^{-2}$)	&	 (M$_{\odot}$)	&	 (cm$^{-3}$)\\	
\hline																																
\hline																																
TeV emission$^a$	&	$-25$ to $-20$	&	7.9	&	1.5	$^{+	0.2	}_{-	0.2	}$	&	9\,000	$^{+	5\,000	}_{-	6\,000	}$	&	30	$^{+	20	}_{-	20	}$	&	 0.3$^{+0.1}_{-0.05}$	&	 1\,000$^{+1\,000}_{-200}$	&	 6$^{+7}_{-2}$ \\	
	&	$-70$ to $-50$	&	7.9	&	3.9	$^{+	1.2	}_{-	0.6	}$	&	30\,000	$^{+	15\,000	}_{-	17\,000	}$	&	90	$^{+	90	}_{-	60	}$	&	 1.3$^{+0.8}_{-0.2}$	&	 6\,000$^{+2\,000}_{-2\,000}$	&	 30$^{+30}_{-10}$ \\	
	&	$-90$ to $-80$	&	7.9	&	1.2	$^{+	0.3	}_{-	0.2	}$	&	8\,000	$^{+	30\,000	}_{-	5\,000	}$	&	20	$^{+	30	}_{-	10	}$	&	 0.2$^{+0.05}_{-0.1}$	&	 700$^{+1\,000}_{-700}$	&	 3$^{+1}_{-3}$ \\	
	&	\multicolumn{2}{c}{Sum of the above}			&	8.1	$^{+	1.7	}_{-	0.9	}$	&	47\,000	$^{+	23\,000	}_{-	27\,000	}$	&	-						&	 1.8$^{+1}_{-0.4}$	&	 8\,000$^{+4\,000}_{-3\,000}$ 	&	\\	
	&	$-110$ to $-100$	&	6.3	&	2.1	$^{+	0.6	}_{-	0.3	}$	&	8\,000	$^{+	3\,000	}_{-	3\,000	}$	&	30	$^{+	50	}_{-	20	}$	&	 0.9$^{+0.1}_{-0.2}$	&	 2\,000$^{+1\,000}_{-1\,000}$	&	 20$^{+20}_{-10}$ \\	
\hline																																
Extended	&	$-25$ to $-20$	&	7.9	&	1.8	&	5\,000	&	50																&	0.3	&	900	&	 6 \\	
X-ray emission	&	$-70$ to $-50$	&	7.9	&	3.2	&	9\,000	&	80																&	0.8	&	 3\,000	&	 20\\	
	&	$-90$ to $-80$	&	7.9	&	0.5	&	1\,000	&	10																&	0.2	&	700	&	 5\\	
	&	$-110$ to $-100$	&	6.3	&	1.4	&	3\,000	&	50																&	0.8	&	 2\,000	&	 20\\	
\hline																																
CXOU\,J171419.8	&	$-25$ to $-20$	&	7.9	&	1.7	&	3\,000	&	60																&	0.4	&	700	&	 10 \\	
$-$383023 	&	$-70$ to $-50$	&	7.9	&	7.1	&	10\,000	&	240																&	1.8	&	 3\,000	&	 60\\	
	&	$-90$ to $-80$	&	7.9	&	2.1	&	3\,000	&	80																&	1.1	&	 2\,000	&	 40 \\	
	&	$-110$ to $-100$	&	6.3	&	3.0	&	3\,000	&	120																&	1.5	&	 1\,000	&	 60\\	
\hline
\end{tabular}
\textit{$^a$Super/subscripts indicate uncertainty propagated from the 1$^{\prime}$ uncertainty in the 68\% containment radius of HESS\,J1714$-$385.\\$^b$Distances are assumed for mass and density calculations, but can be scaled from the assumed distance, $d_0$, to a different distance, $d$, by multiplying by $\left(d/d_0\right)^2$ and $\left(d/d_0\right)^{-1}$ for mass and density, respectively, if further distance constraints arise in the future.}
\end{table*}
\end{landscape}

\begin{table*}
\centering
\caption{Same as Table\,\ref{Table:GasParams}, but parameters are from HI absorption analyses (and emission analyse where specified) and showing atomic (in contrast to molecular) H column density. \label{Table:GasParamsHI}}
\begin{tabular}{l|l|l|r|c|l}
\hline
Region	&Velocity	& Assumed	& \multicolumn{3}{c}{HI} 	\\
		&Range		& Distance, $d_0$	& $\overline{N_{H}}$	&  $M$	&	$\overline{n_{H}}$ \\
		&(km\,s$^{-1}$)&(kpc)		& (10$^{21}$\,cm$^{-2}$)& (M$_{\odot}$)& (cm$^{-3}$) \\
\hline		
\hline
TeV emission			&$^a-70$ to $-50$ & 7.9	& 5.6$^{+0.1}_{-0.1}$	& 	9\,000$^{+3000}_{-6000}$		& 	130$^{+100}_{-20}$ \\
&	5	to	8			&	-	&	0.3$^{+0.1}_{-0.05}$		&	-						&	-	\\
&	0	to	2			&	-	&	0.3$^{+0.1}_{-0.05}$		&	-						&	-	\\
&	$-$13	to	$-$7	&	1.5	&	1.9$^{+0.9}_{-0.3}$			&	100$^{+100}_{-0}$		&	200$^{+200}_{-100}$	\\
&	$-$24	to	$-$18	&	7.9	&	2.8$^{+1.8}_{-0.2}$			&	6\,000$^{+3000}_{-500}$	&	50$^{+50}_{-10}$	\\
&	$-$74	to	$-$71	&	7.9	&	0.1$^{+0.05}_{-0.05}$		&	300$^{+100}_{-100}$		&	3$^{+1}_{-1}$	\\
&	$-$90	to	$-$86	&	7.9	&	0.2$^{+0.0}_{-0.05}$		&	400$^{+200}_{-200}$		&	4$^{+1}_{-1}$	\\
&	$-$107	to	$-$103	&	6.3	&	0.2$^{+0.1}_{-0.0}$			&	300$^{+200}_{-100}$		&	6$^{+4}_{-1}$	\\
&\multicolumn{2}{c}{Absorption sum} & 5.8$^{+3.0}_{-0.4}$		&	7\,000$^{+3000}_{-1000}$&	-	\\
\hline
Extended X-ray emission		&$^a-70$ to $-50$	& 7.9	& 4.6	& 5\,000	& 100 \\
&	5	to	8			&	-	&	0.3	&	-	&	-	\\
&	0	to	2			&	-	&	0.4	&	-	&	-	\\
&	$-$13	to	$-$7	&	1.5	&	1.0	&	50	&	100	\\
&	$-$23	to	$-$18	&	7.9	&	1.2	&	2\,000	&	30	\\
&	$-$79	to	$-$76	&	7.9	&	0.2	&	300	&	4	\\
&	$-$92	to	$-$87	&	7.9	&	0.3	&	500	&	8	\\
&	$-$107	to	$-$104	&	6.3	&	0.6	&	500	&	20	\\
&\multicolumn{2}{c}{Absorption sum} & 4.0 &-&-	\\
\hline				
CXOU\,J171419.8$-$383023	&$^a-70$ to $-50$	& 7.9	& 5.9	& 4\,000	& 200\\
&	5	to	8	&	-	&	0.2	&	-	&	-	\\
&	0	to	2	&	-	&	0.4	&	-	&	-	\\
&	$-$13	to	$-$7	&	1.5	&	1.3	&	40	&	200	\\
&	$-$23	to	$-$18	&	7.9	&	2.2	&	2\,000	&	70	\\
&	$-$79	to	$-$76	&	7.9	&	0.3	&	200	&	10	\\
&	$-$92	to	$-$87	&	7.9	&	0.5	&	400	&	20	\\
&	$-$107	to	$-$104	&	6.3	&	0.6	&	300	&	30	\\
&\multicolumn{2}{c}{Absorption sum} & 5.5 &-&-	\\
\hline
\end{tabular}
\\ \textit{$^a$From HI emission analyses.}
\end{table*}

\begin{table*}
\caption{Gas parameters towards locations with ON-OFF switched deep CS(1-0) observations (see Figure\,\ref{fig:CTB1}). CS(1-0)-derived average column density, $N_{H_2}$, mass, $M$, and average density, $n_{H_2}$, within selected regions are shown for significant CS(1-0) emission lines (see Table\,\ref{Table:DeepPointings}). All calculated values have an associated statistical uncertainty of $\sim$15\%. \label{Table:GasParamsPointings}}
\begin{tabular}{l|l|l|r|c|l}
\hline
Location&Velocity		& Assumed	& \multicolumn{3}{c}{CS(1-0)} \\
		&				& Distance	& $\Nh$ 				&	$M$			& \nh \\
		&(km\,s$^{-1}$)	&\multicolumn{2}{c}{(kpc)~~~~(10$^{21}$\,cm$^{-2}$)}& (M$_{\odot}$)	& (cm$^{-3}$)\\
\hline
\hline
1		&$-$73.4	& 7.9 	& 49	& 1\,500& 7$\times$10$^{3}$\\
\hline
3		&$-$105.1	& 6.3	& 39	& 800	& 7$\times$10$^{3}$\\
				&$-$88.0	& 7.9	& 6.8	& 200	& 1$\times$10$^{3}$\\
				&$-$57.6 	& 7.9	& 75	& 2\,300& 1$\times$10$^{4}$\\
				&$-$22.4  	& 7.9	& 19	& 600	& 3$\times$10$^{3}$\\
\hline
4		&$-$61.9	& 7.9	& 17	& 500	& 3$\times$10$^{3}$\\
				&$-$21.9	& 7.9	& 3.9	& 100	& 600\\
\hline
5		&$-$23.0	& 7.9	& 11	& 400	& 1$\times$10$^{3}$\\
\hline
6		&$-$63.9	& 7.9	& 34	& 1\,100& 5$\times$10$^{3}$\\
\hline								
7		&$-$103.9	& 6.3	& 8.9	& 170 	& 2$\times$10$^{3}$\\
				&$-$89.2	& 7.9	& 1.2	& 370	& 2$\times$10$^{3}$\\
				&$-$64.2	& 7.9	& 50	& 1\,500& 7$\times$10$^{3}$\\
\hline								
\end{tabular}
\end{table*}

\section{Discussion}
\subsection{Previous X-ray Absorption Studies}\label{sec:X-ray}
CTB\,37A exhibits extended thermal X-ray emission, possibly from a SNR shock-gas interaction, and a more compact non-thermal source, CXOU\,J171419.8$-$383023 (Chandra beam FWHM $\sim$1$^{\prime\prime}$), which may be a PWN \citep{Aharonian:2008}. \citet{Aharonian:2008} used a spectral-fitting analysis technique to estimate the level of X-ray-absorption, and hence the column density towards these two sources. We compare column densities estimated from CO(2-1), CS(1-0) and HI data to column densities calculated from X-ray absorption. The column densities and masses of significant molecular and atomic gas foreground to the CTB\,37A extended X-ray emission and CXOU\,J171419.8$-$383023 are presented in Tables\,\ref{Table:GasParams} and \ref{Table:GasParamsHI}. The X-ray emission regions are highlighted in Figures\,\ref{fig:CSpics}a, c and d.

Towards the extended thermal X-ray emission region, we calculated a lower limit for proton column density ($\overline{N_{H}}+$2\Nh, where $\overline{N_{H_2}}$ is the average of that derived for CO(2-1) and CS(1-0)) of $\sim$9$\times$10$^{21}$\,cm$^{-2}$. \citet{Aharonian:2008} calculated an X-ray absorption proton column density of $\sim$3$\times$10$^{22}$\,cm$^{-2}$ consistent with our lower limit and work by \citet{Sezer:2011}.

Similar to towards the extended thermal X-ray emission region, towards CXOU\,J171419.8$-$383023, we calculated a lower limit on proton column density foreground to CTB\,37A ($\overline{N_{H}}+$2\Nh $\sim$1.5$\times$10$^{22}$\,cm$^{-2}$) that was consistent with the X-ray absorption proton column density of $\sim$6$\times$10$^{22}$\,cm$^{-2}$.

We note that the molecular gas fractions (H$_2$/[H+H$_2$]) for these two regions are $f_{H2}\sim$0.38 and $\sim$0.47. \citet{Liszt:2010} investigated the atomic gas fraction from a sample of sources using HCO$^+$ emission, CO emission and HI absorption and found that the molecular gas fraction was $f_{H2}\sim$0.35, i.e. $\sim$2$\times$ more atomic H atoms exist than H$_2$ molecules. The authors noted the results of other methods that put $f_{H2}\gtrsim$0.25 \citep{Bohlin:1978} and $f_{H2}\thickapprox$0.40-0.45 \citep{Liszt:2002}. Our calculated molecular gas fractions (0.38 and 0.47) are consistent with previous analyses, adding confidence to the accuracy of our column density estimations. There is, however, a factor $\sim$3-4 difference between our lower limits on column density and the X-ray absorption column densities. This may be due to gas components being overlooked by our observation or analysis.


\subsection{Towards HESS\,J1714$-$385}\label{sec:tev}
\begin{table*}
\centering
\caption{Atomic mass (from HI analyses), molecular mass (average of CO and CS analyses) are summed and multiplied by 1.33 to account for a 25\% He component. The ranges displayed reflect the uncertainty in the 68\% containment radius of HESS\,J1714-385. An estimate of CR enhancement factor above the solar system value ($>$1\,TeV), $k_{CR}$, assuming that all mass coincident the best-fit region of HESS\,J1714-385 acts as as CR target material, is also displayed.\label{tab:CRenhance}}
\begin{tabular}{l|l|l|r|c|l|l}
\hline
Region		Velocity	&	 Assumed	&	 Atomic mass 			&	 Molecular mass 			&	 Total mass			&	 $\sim k_{CR}~^a$			\\
		Range	&	 Distance, $d_0$	&	 			&	  			&	(25\% He)			&				\\
		(km\,s$^{-1}$)	&	(kpc)	&	 (M$_{\odot}$) 			&	 (M$_{\odot}$) 			&	 (M$_{\odot}$)			&				\\
		\hline																			
		\hline																			
TeV emission		$-25$ to $-20$	&	7.9	&	5\,500	\,-\,	9\,000	&	800	\,-\,	14\,000	&	8\,400	\,-\,	31\,000	&	320	\,-\,	1\,200	\\
		$-70$ to $-50$	&	7.9	&	3\,000	\,-\,	12\,000	&	4\,000	\,-\,	45\,000	&	9\,300	\,-\,	76\,000	&	130	\,-\,	1\,100	\\
		$-90$ to $-80$	&	7.9	&	200	\,-\,	600	&	0	\,-\,	11\,000	&	300	\,-\,	15\,000	&	660	\,-\,	33\,000	\\
		\multicolumn{2}{c}{Sum of the above}	&	8\,700	\,-\,	22\,000	&	4\,800	\,-\,	70\,000	&	18\,000	\,-\,	120\,000	&	80	\,-\,	550	\\
		$-110$ to $-100$	&	6.3	&	200	\,-\,	500	&	1\,000	\,-\,	11\,000	&	1\,600	\,-\,	15\,000	&	420	\,-\,	6\,200	\\
\hline
\end{tabular}
\textit{\\$^a$CR enhancement factor is effectively independent of the assumed distance (as distance cancels in Equation\,\ref{equ:Aharonian} when the distance assumptions of our mass calculations are accounted for.)}
\end{table*}

If the observed TeV emission is produced by hadronic processes (i.e. CR protons or nuclei colliding with matter to produce neutral pions that decay into gamma-ray photons), constraining the mass of matter may help to constrain the CR density in the region. \citet{Aharonian:1991} derived a relation to predict the flux of gamma-rays from the mass of CR-target material, assuming an $E^{-1.6}$ integral power law spectrum. The expected gamma ray flux above $E_{\gamma}$ is,
\begin{equation}
\label{equ:Aharonian}
F\left(\geq E_{\gamma} \right) = 2.85\times 10^{-13} E_{TeV}^{-1.6} \left( \frac{M_5}{d_{kpc} ^2 }\right) k_{CR}~~~~\textrm{cm$^{-2}$s$^{-1}$}
\end{equation} where $M_5$ is the gas mass in units of 10$^5$\,M$_{\odot}$, $d_{kpc}$ is the distance in units of kpc, $k_{CR}$ is the CR enhancement factor above that observed at Earth and $E_{\gamma}>$1\,TeV. The gamma-ray flux above 1\,TeV towards CTB\,37A is $F(>1\,\textrm{TeV})\sim$6.7$\times$10$^{-13}$\,cm$^{-2}$s$^{-1}$ \citep{Aharonian:2008} and the masses within the HESS best fit region (68\% containment of the gamma-ray source) were calculated for given velocity ranges (see Tables \ref{Table:GasParams} and \ref{Table:GasParamsHI}). The cosmic ray enhancement factor, $k_{CR}$, could thus be calculated for several velocity ranges. These are displayed in Table\,\ref{tab:CRenhance}. 

CR enhancement factor estimates (above 1\,TeV) for the gas at \vlsr$\sim -$22, $-$85 and $-$110\,km\,s$^{-1}$ range between 320 and 33\,000, whereas the CR enhancement factor esimate for the gas at \vlsr$\sim-$60\,km\,s$^{-1}$, the gas most likely to be associated with CTB\,37A, generally spans a range of smaller values, 130\,-\,1\,100. We assume that the gas at $\vlsr\sim-$60\,km\,s$^{-1}$ is associated with CTB\,37A, while the gas components at $\vlsr\sim-$22 and $-$85\,km\,s$^{-1}$ may or may not additionally be associated, giving a range of possible CR-enhancements in the CTB\,37A region of $\sim$80\,-\,1\,100$\times$ (see Table\,\ref{tab:CRenhance}) that observed in the local solar neighbourhood, assuming a hadronic scenario. 

In a hadronic scenario for gamma-ray production, matter is expected to overlap with gamma-ray emission on a large-scale, because the TeV gamma-ray flux is proportional to the product of mass, $M_5$, and cosmic ray enhancement, $k_{CR}$. If one assumes a uniform CR-density throughout the CTB\,37A region, a good correlation between matter density and hadronic gamma-ray emission might be expected, which, from CO(2-1) and CS(1-0) emission (see Figure\,\ref{fig:CSpics}), may be the case for line of sight velocity, $\sim -$85\,km\,s$^{-1}$, but less likely to be so for the line of sight velocities, $\sim -$22\,km\,s$^{-1}$ or $\sim -$110\,km\,s$^{-1}$. Gamma-ray flux variation can also reflect variation in cosmic ray enhancement, such that a region can have matter existing outside of the region indicated by gamma-ray emission contours (see Figure\,\ref{fig:CSpics}) while still being consistent with a hadronic emission scenario. With this in mind, the line of sight velocity $\sim -$60\,km\,s$^{-1}$, the suspected location of CTB\,37A, is consistent with a hadronic scenario. When we summed multiple emission components, i.e. the \vlsr$\sim -$22\,km\,s$^{-1}$, $-$60\,km\,s$^{-1}$ and $-$85\,km\,s$^{-1}$ gas, the \vlsr$\sim -$60\,km\,s$^{-1}$ component is dominant, such that only minimal improvement in molecular gas-gamma ray overlap was observed. It seems that from studies of the overlap between molecular gas and gamma-ray emission alone, no clear conclusion about which gas clumps correspond to CTB\,37A can be drawn.

The proportion of gas that may be acting as target material towards the HESS\,J1714$-$385 region is unknown, so CR enhancement estimates in Table\,\ref{tab:CRenhance} may be considered as lower than plausibly expected. Conversely, a component of dark gas (where carbon exists in an atomic/ionic form and hydrogen exists in molecular form) towards the HESS\,J1714$-$385 region is not taken into account (and may indeed exceed 30\% according to \citeauthor{Wolfire:2010}, \citeyear{Wolfire:2010}), making the CR enhancement estimates higher than plausibly expected. Due to these two unconstrained effects, CR enhancement estimates in Table\,\ref{tab:CRenhance} should be considered with some caution.

We note that towards the HESS\,J1714-385 TeV emission region, a calculation of the kinetic energy \citep{Arikawa:1999,Nicholas:2011} implied by observed CS(1-0) emission at $\vlsr\sim-$60\,km\,s$^{-1}$, $E\sim \frac{1}{2} M \fwhm ^2$, is $\sim$0.4\,-\,4$\times$10$^{50}$\,erg. This is a significant fraction (4\,-\,40\%) of the assumed SNR ejecta kinetic energy of $\sim$10$^{51}$\,erg, so may in fact be injected by CTB\,37A. The CS(1-0) emission at $\vlsr\sim-$22\,km\,s$^{-1}$ is also broad, and implies a kinetic energy of $\sim$0.7-3$\times$10$^{49}$\,erg, $\sim$0.7-3\% of the assumed ejecta energy. Both the \vlsr$\sim -$22 and $-$60\,kms$^{-1}$ gas clouds have velocity dispersions consistent with a SNR-shock interaction.


\subsection{The Age of CTB\,37A}\label{sec:Age}
The age of CTB\,37A has been considered by \citet{Sezer:2011} using the assumption of full-ionisation equilibrium to calculate a plasma age of $\sim$3$\times$10$^4 \sqrt{f}$, where $f$ is the plasma filling-factor within a spherical region. With the benefit of the latest distance analyses by TL2012, this estimate may be scaled upwards by $\sim$20-80\%, to account for a nearer distance to CTB\,37A (6.3-9.5\,kpc).

To complement the work of \citet{Sezer:2011}, similarities of CTB\,37A with other SNRs cannot be overlooked. Referring to \citet{Green:1997}, which summarises SNRs with detected associated OH maser emission, and cross-checking these with SNR catalogues \citep{Guseinov:2003,Guseinov:2004a,Guseinov:2004b}, a short list of OH maser-associated SNRs with estimated ages can be compiled: W28 ($\sim$10$^4$\,yr), W44 ($\sim$10$^4$\,yr), W51c ($\sim$10$^4$\,yr), IC\,443 ($\sim$10$^3$\,yr), CTB\,33 ($\leq$10$^5$\,yr) and G347$+$0.2 ($\sim$10$^3$\,yr). Of these, the first 4 SNRs have diameters of $\sim$15-35\,pc, similar to CTB\,37A which has dimensions of $\sim$0.12$^{\circ}\times$0.23$^{\circ}$ and an average diameter of $\sim$24$\pm$5\,pc at a distance of 7.9$\pm$1.6. Based purely on these similarities, the CTB\,37A age may be of the order of 10$^3$-10$^4$\,yr.

Further support for a $\sim$10$^3$-10$^4$\,yr age for CTB\,37A comes from assuming a standard Sedov-Taylor time-scaling of the shock-radius (here taken from \citet{Caprioli:2009}, Equation\,29). For an ejecta with energy 10$^{51}$\,erg expanding into a medium of number density 10\,cm$^{-3}$, similar to towards HESS\,J1714-385 (although the shock radius is only weakly dependent on density, being proportional to $n^{-1/5}$), the shock radius reaches $\sim$24$\pm$5\,pc after $\sim$0.6$^{+0.4}_{-0.3}\times$10$^4$\,yr. Certainly, the old age is consistent with the large fraction of energy possibly injected into the \vlsr$\sim - 60$\,km\,s$^{-1}$ cloud by CTB\,37A (4-40\%, \S\,\ref{sec:tev}), assuming an ejecta energy of 10$^{51}$\,erg.

Finally, \citet{Aharonian:1996} model the diffusion of CRs away from a continuous accelerator and plot the energy-dependent CR-enhancement factor for different SNR ages. We assume that slow CR diffusion, like that seen towards the similar SNR, W28 (see eg. \citeauthor{Aharonian:2008w28}, \citeyear{Aharonian:2008w28} and \citeauthor{Gabici:2010}, \citeyear{Gabici:2010}), applies to the CTB\,37A region. On referring to Figure\,2b in \citet{Aharonian:1996}, which corresponds to a galactic CR diffusion coefficient (parameterised from a CR energy of 10\,GeV) of 10$^{26}$\,cm$^2$s$^{-1}$ and a source CR spectrum of spectral index $\gamma=-$2.2, a CR-enhancement of $\sim$80-300 at TeV energies corresponds to an age of 10$^4$-10$^5$\,yr (at a distance of 10\,pc from the CR release point). The range of the models considered by \citet{Aharonian:1996} does not suggest CR enhancement factors above $\sim$300. We note that the CR enhancement factor resulting from a normal (in contrast to `slow') galactic diffusion coefficient of 10$^{28}$\,cm$^2$s$^{-1}$ is substantially lower (2-3$\times$ the local value), inconsistent with the calculated CR enhancement range, but we favour the W28-region CR diffusion speed due to the physical similarities with CTB\,37A (see above).

The range of age values suggested by this CR enhancement argument (10$^4$-10$^5$\,yr) spans larger values than those suggested by the previous 3 methods (the plasma age, comparison with similar remnants and the Sedov-Taylor scaling relation). If a leptonic component of gamma-ray emission was present, CR enhancement would be less than that calculated here. It also follows that the age would be overestimated by our CR enhancement factor argument.

Through this CR-enhancement factor argument, the simplistic shock-radius model, and the aforementioned similarities to other SNRs, we find that the age of CTB\,37A is consistent with the previous result of \citet{Sezer:2011} (for $f\sim$1), of the order $\sim$10$^4$\,yr.

\section{Conclusions}
Using CO(2-1), CS(1-0) and HI spectra, we conducted an investigation of the atomic and molecular gas towards the CTB\,37A region, which has signatures of a shock-gas interaction (OH masers, thermal X-rays) and is a source of TeV gamma rays, possibly hadronic in origin.


CS(1-0) observations identified new dense gas components associated with 4 of the 6 observed locations that exhibit 1720\,MHz OH maser emission.

CO(2-1), CS(1-0) and HI emission, and HI absorption allowed the estimation of lower limits of column density towards regions of interest. Column density lower limits towards X-ray emission regions are consistent with X-ray absorption measurement-derived column densities.

CO(2-1), CS(1-0) and HI-derived mass estimates for specific gas components towards the CTB\,37A region allowed an investigation of the CR hadron-target mass available. Assuming that all measured mass (and an assumed additional 25\% He mass) in the HESS gamma-ray emission region is potential CR target material in a hadronic scenario for TeV gamma-ray emission, we estimate a CR density of $\sim$80\,-\,1\,100$\times$ that seen at Earth.

Based on a comparison of CTB\,37A to other SNRs associated with OH-maser emission, Sedov-Taylor phase scaling of the CTB\,37A shock radius, and an examination of the expected change in CR-enhancement factor with time, we estimate the CTB\,37A age to be consistent with previous estimates of $\sim$10$^4$\,yr.

High resolution ($\sim$1$^{\prime}$), high energy (GeV-TeV) measurements, such as what may be expected from future ground-based gamma-ray telescopes \citep{CTA:2011}, would allow a more detailed investigation of the nature of the high energy spectrum towards CTB\,37A, especially with regard to the spatial correspondence with molecular gas. 

\section{Acknowledgments}
This work was supported by an Australian Research Council grant (DP0662810, 1096533). The Mopra Telescope is part of the Australia Telescope and is funded by the Commonwealth of Australia for operation as a National Facility managed by the CSIRO. The University of New South Wales Mopra Spectrometer Digital Filter Bank used for these Mopra observations was provided with support from the Australian Research Council, together with the University of New South Wales, University of Sydney, Monash University and the CSIRO.

We thank Naomi McClure-Griffiths for feedback regarding our HI analysis and our anonymous referee for his/her detailed constructive criticism of our manuscript.

\appendix

\section{Infrared Emission towards CTB\,37A}\label{sec:infra}
Evidence for shock-interactions are also seen at infrared wavelengths. \citet{Reynoso:2000} noted the presence of shocked, heated dust, IRAS\,17111$-$3824, towards a region of overlap between CTB\,37A and G348.5$-$0.0. \citet{Reach:2006} later discovered several patches and filaments of 4.5\,$\mu$m emission towards and around Locations 5 and 6, possibly indicating shocked H$_2$ gas. The authors also note further evidence for shocked gas at 5.8-8\,$\mu$m towards the CTB\,37A-G348.5$-$0.0 overlap region.

\begin{figure*}
\centering
\includegraphics[width=0.99\textwidth]{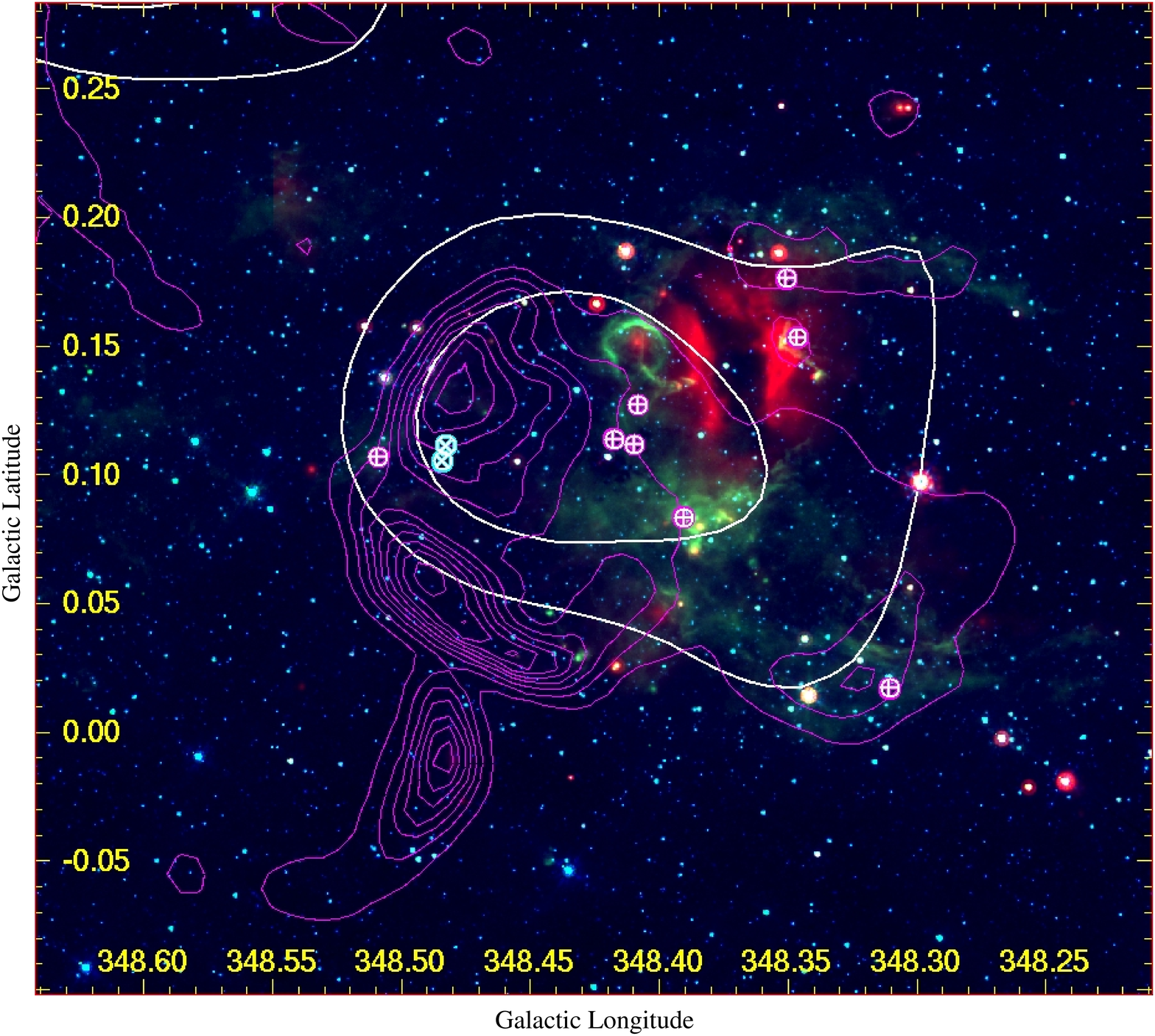}
\caption{Spitzer 24\,$\mu$m, 8\,$\mu$m and 5.8\,$\mu$m emission are shown in red, green and blue, respectively. White contours show HESS 80 and 100 excess count gamma-ray emission, and magenta contours show Molonglo 843\,MHz radio continuum emission. Crosses represent both the $\sim-$65\,km\,s$^{-1}$ (pink,$+$) and $\sim-$25\,km\,s$^{-1}$ (cyan,$\times$) OH maser populations \citep{Frail:1996}. \label{fig:CTB_infrared}}
\end{figure*}

Figure\,\ref{fig:CTB_infrared} is an image of 24, 8 and 5.8\,$\mu$m infrared emission. Millimetre star-formation indicators (HC$_3$N, CH$_3$OH, SiO, Figure\,\ref{fig:CTB_Species}) are seen towards Location\,3 (\vlsr$\sim -$110\,km\,s$^{-1}$), but no clear association is seen at infrared wavelengths. Faint 8\,$\mu$m emission may indicate warm dust, but this extends over a large region of the centre of CTB\,37A and is most intense in the region surrounding Location\,4. This characteristic, however, does not rule out the notion that warm dust may be responsible for the millimetre molecular emission towards Location\,3. 

North of Location\,3, a shell of 8\,$\mu$m emission 0.02$-$0.03$^{\circ}$ in diameter, can be seen surrounding a compact 24\,$\mu$m source. This is likely to be a wind-blown bubble surrounding a high-mass O or B star (object S5 from \citeauthor{Churchwell:2006}, \citeyear{Churchwell:2006}), unrelated to CTB\,37A and HESS\,J1714$-$385.

Emission at 24\,$\mu$m surrounding ($l$,$b$)$\sim$(348.37,0.14) clearly shows hot dust obscured by a infrared-dark absorber, likely a cold foreground cloud. CO(2-1) emission at \vlsr$\sim -$10\,km\,s$^{-1}$ is observed towards this region, but the CO-traced gas is too uniform across CTB\,37A's north-west to explain the localised `finger' of infrared-absorption. Observations of the cold, dense gas tracer NH$_3$(1,1) were employed in an attempt to aid in constraining the absorber's location, but we found no NH$_3$(1,1) emission towards this location in the low-exposure HOPS galactic plane data \citep{Walsh:2008} and a follow-up high-exposure (\Trms$\sim$0.1\,K\,ch$^{-1}$) deep pointing revealed no NH$_3$(1,1) emission either.

\section{Additional 7\,mm Images Towards CTB\,37A}\label{sec:addimages}
\begin{figure*}
\centering
\includegraphics[width=0.49\textwidth]{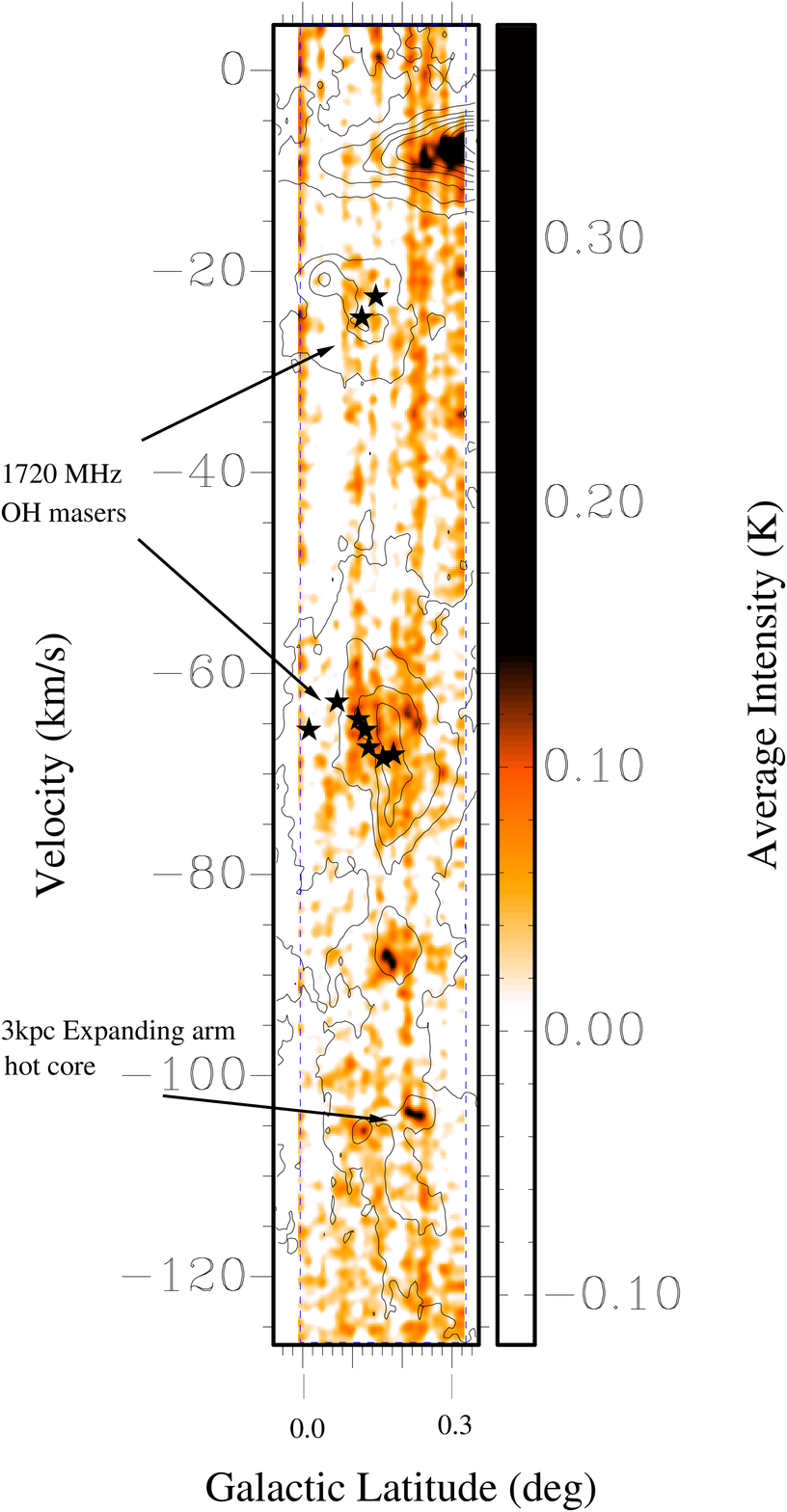}
\caption{Position-velocity image of Mopra CS(1-0) emission towards the CTB\,37A region. Nanten CO(2-1) emission contours towards the CTB\,37A region are overlaid. 
Black stars indicate 1720\,MHz OH maser locations and the dashed lines indicate the extent of the 7\,mm mapping campaign for CS(1-0) emission. \label{fig:PVplot_lat}}
\end{figure*}

Figure\,\ref{fig:PVplot_lat} is a position-velocity image towards the CTB\,37A region. Similar to Figure\,\ref{fig:PVplot}, several clouds are visible in CO(2-1) at approximate line-of-sight reference velocities, $\vlsr\sim$ $-$10, $-$20, $-$60 to $-$75, $-$90 and $-$105\,km\,s$^{-1}$. Corresponding CS(1-0) emission is observed towards CO(2-1)-traced clouds at approximate line-of-sight reference velocities, $\vlsr\sim$ $-$10, $-$60 to $-$75, $-$90 and $-$105\,km\,s$^{-1}$, indicating dense gas at these locations.

\begin{figure*}
\centering
\includegraphics[width=0.49\textwidth]{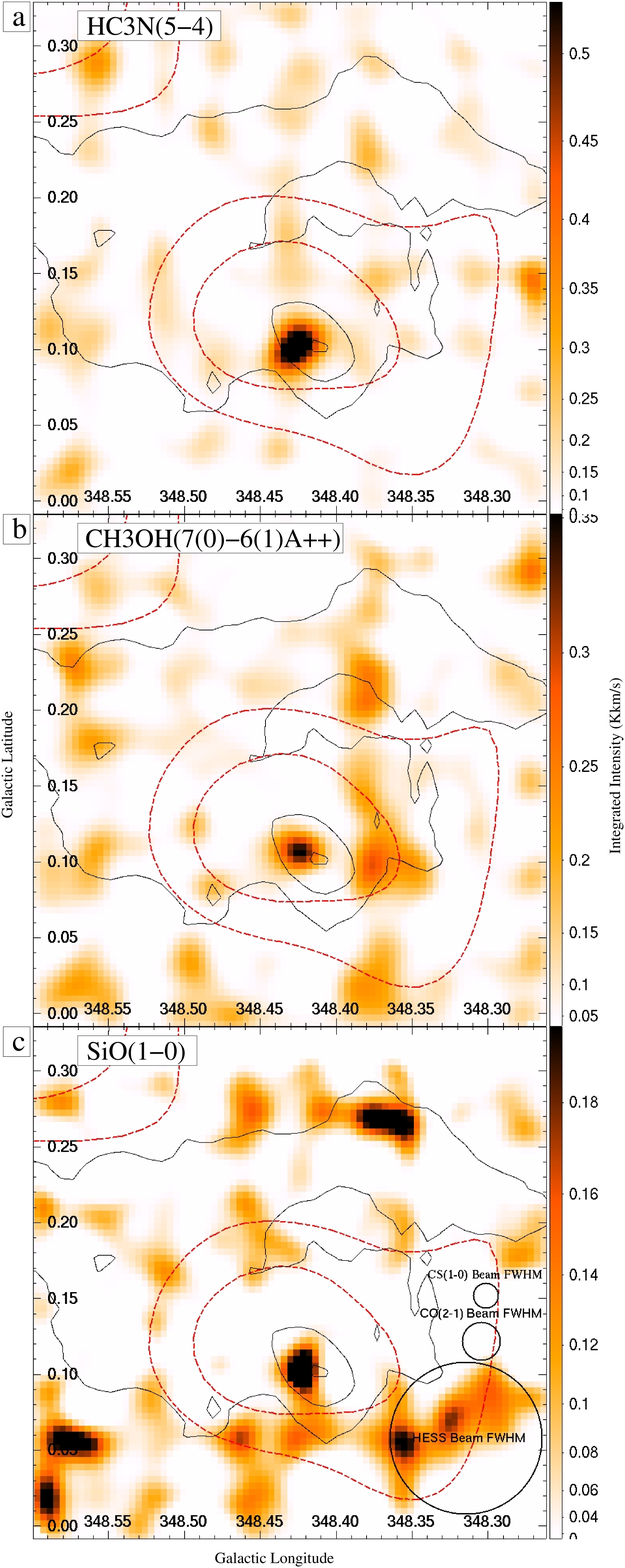}
\caption{Various spectral line emissions integrated between $\vlsr =-$108 and $-$102\,km\,s$^{-1}$. Black contours indicate CO(2-1) integrated intensity levels of 10, 20 and 30\,K\,km\,s$^{-1}$ and red dashed contours indicate HESS 80 and 100 excess count gamma-ray emission. See \S\,\ref{sec:-110} for details. \label{fig:CTB_Species}}
\end{figure*}

Figure\,\ref{fig:CTB_Species}a/b/c are three images of emission from the HC$_3$N, CH$_3$OH and SiO molecules. The narrow spectral line profile and correspondence with a molecular core seen in CO(2-1) and CS(1-0) emission, suggest possible star-formation activity.

\end{document}